\documentclass[review,12pt]{elsarticle}

\usepackage{lineno,hyperref}
\modulolinenumbers[5]

\usepackage{etex}
\usepackage{amsmath}
\usepackage{amssymb}
\usepackage{bm}
\usepackage{color}
\usepackage{url}
\usepackage{wrapfig,lipsum,booktabs}
\usepackage{graphicx}
\usepackage{xcolor}
\usepackage[all]{xy}

\usepackage{pgf}
\usepackage{tikz}
\usetikzlibrary{arrows,automata,positioning}
\usetikzlibrary{shapes,arrows,positioning}
\tikzset{
    >=stealth',
    punkt/.style={
           rectangle,
           rounded corners,
           draw=black, very thick,
           text width=6.5em,
           minimum height=2em,
           text centered},
    pil/.style={
           ->,
           thick,
           shorten <=2pt,
           shorten >=2pt,}
}

\definecolor{shadecolor}{rgb}{0.2,0.9,0.2}
\definecolor{green}{rgb}{0,0.4,0}

\usepackage{adjustbox}
\usepackage{hyperref}
\hypersetup{
    unicode=false,          
    pdftoolbar=true,        
    pdfmenubar=true,        
    pdffitwindow=false,     
    pdfstartview={FitH},    
    pdftitle={GPs for EO},    
    pdfauthor={Camps-Valls},     
    pdfsubject={},   
    pdfcreator={},   
    pdfproducer={}, 
    pdfkeywords={Forward}{inverse}{modelling}{Gaussian process}{kernels}, 
    pdfnewwindow=true,      
    colorlinks=true,       
    linkcolor={blue},          
    citecolor=blue,        
    filecolor=blue,      
    urlcolor=blue           
}

\newcommand{\Real}{{\mathbb R}}
\newcommand{\balpha}{\boldsymbol{\alpha}}

\newcommand{\bomega}{\boldsymbol{\omega}}
\newcommand{\btheta}{\boldsymbol{\theta}}

\newcommand{\x}{\mathbf{x}}
\newcommand{\y}{\mathbf{y}}

\newcommand{\I}{\mathbf{I}}

\newcommand{\X}{\mathbf{X}}
\newcommand{\Y}{\mathbf{Y}}

\newcommand{\bk}{\mathbf{k}}
\newcommand{\bK}{\mathbf{K}}

\newcommand{\dataset}{{\mathcal D}}
\newcommand{\Normal}[0]{\mathcal{N}}

\newcommand{\mse}{\textrm{MSE}}
\newcommand{\rmse}{\textrm{NMSE}}
\newcommand{\lai}{\textrm{LAI}}
\newcommand{\fapar}{\textrm{fAPAR}}
\newcommand{\thetavec}{\boldsymbol{\theta}}
\newcommand{\muvec}{\boldsymbol{\mu}}

\newcommand{\diff}{\textnormal{d}}

\newcommand{\linOpSpecific}[3]{\textrm{L}_{#1}[#2]\left\{#3\right\}}

\renewcommand{\vec}[1]{\mathbf{#1}}

\begin{document}

\begin{frontmatter}

\title{Physics-Aware Gaussian Processes in Remote Sensing\footnote{{\bf Preprint. Paper published in Applied Soft Computing Volume 68, July 2018, Pages 69-82. https://doi.org/10.1016/j.asoc.2018.03.021}}}

\author[IPL]{Gustau Camps-Valls}
\author[IPL]{Luca Martino}
\author[IPL]{Daniel H. Svendsen}
\author[TERMO]{Manuel Campos-Taberner}
\author[IPL]{Jordi Mu{\~n}oz-Mar\'i}
\author[IPL]{Valero Laparra}
\author[UPM]{David Luengo}
\author[TERMO]{Javier Garc\'ia-Haro}

\address[IPL]{Image Processing Laboratory (IPL), Universitat de Val\`encia, Spain}
\address[TERMO]{Faculty of Physics, Universitat de Val\`encia, Spain}
\address[UPM]{Signal Processing and Communications Dep., Univ. Polit\'ecnica de Madrid, Spain}

\begin{abstract}
Earth observation from satellite sensory data poses challenging problems, where machine learning is currently a key player. In recent years, Gaussian Process (GP) regression have excelled in biophysical parameter estimation tasks from airborne and satellite observations. GP regression is based on solid Bayesian statistics, and generally yields efficient and accurate parameter estimates. However, GPs are typically used for inverse modeling based on concurrent observations and {\em in situ} measurements only. Very often a {\em forward model} encoding the well-understood physical relations between the state vector and the radiance observations is available though and could be useful to improve predictions and understanding. 
In this work, we review three GP models that respect and learn the physics of the underlying processes in the context of both {\em forward and inverse modeling}. After reviewing the traditional application of GPs for parameter retrieval, we introduce a Joint GP (JGP) model that combines {\em in situ} measurements and simulated data in a single GP model. Then, we present a latent force model (LFM) for GP modeling that encodes ordinary differential equations to blend data-driven modeling and physical constraints of the system governing equations. The LFM performs multi-output regression, adapts to the signal characteristics, is able to cope with missing data in the time series, and provides explicit latent functions that allow system analysis and evaluation. Finally, we present an Automatic Gaussian Process Emulator (AGAPE) that approximates the forward physical model using concepts from Bayesian optimization and at the same time builds an optimally compact look-up-table for inversion. We give empirical evidence of the performance of these models through illustrative examples of vegetation monitoring and atmospheric modeling.
\end{abstract}


\end{frontmatter}

\clearpage
\tableofcontents
\newpage

\section{Introduction}

Solving inverse problems is a recurrent topic of research in Engineering and Physics in general, and in Remote Sensing and Earth Observation (EO) in particular.
A very relevant inverse problem is that of estimating vegetation properties from remotely sensed images. Accurate inverse models help to determine the phenological stage and health status (e.g., development, productivity, stress) of crops and forests~\cite{Hilker08}, which has important societal, environmental and economical implications. Leaf area index (LAI) defined as half the total intercepting leaf area per unit ground surface area~\cite{CHENandBLACK}, leaf chlorophyll content ($Chl$), fraction of absorbed photosynthetically active radiation (fAPAR), and fractional vegetation cover (FVC) are among the most important vegetation parameters to retrieve from space observations~\cite{Whittaker75,Lichtenthaler87}. 

\begin{figure}[t!]
\begin{center}
\begin{tikzpicture}[->,>=stealth',scale=.85,transform shape,node distance=4cm,thick]
  \tikzstyle{every state}=[fill=green!30,draw=none,text=black]
  \node[state] (RET)  [fill=red!60] {Retrieval $f(\x,\theta)$};
  \node[state] (RTM)  [above of=RET,fill=red!60] {RTM $g(\y,\omega)$};
  \node[state] (OBS)  [right of=RTM,fill=red!80,yshift=-2cm] {Observations $\x$};
  \node[state] (VARS) [left of=RTM,fill=red!80,yshift=-2cm]  {Variables $\y$};
  \path[solid, bend left=25, thick, color=red] (VARS) edge node {} (RTM);
  \path[solid, bend left=25, thick, color=red] (RTM)  edge node[yshift=+0.5cm,xshift=+1cm] {forward problem} (OBS);
  \path[dashed, bend left=25, thick, color=red] (OBS)   edge node {} (RET);
  \path[dashed, bend left=25, thick, color=red] (RET)   edge node[yshift=-0.5cm,xshift=-1cm] {inverse problem} (VARS);
\end{tikzpicture}
\end{center}
\vspace{-0.25cm}
\caption{Forward (solid lines) and inverse (dashed lines) problems in remote sensing.}
\label{forward_inverse}
\end{figure}
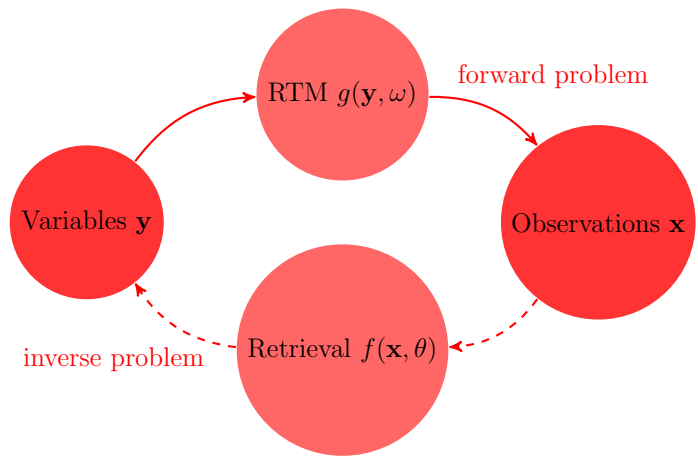

In general, {physical models implement the laws of Physics and allow us to compute the observation values given a state and a model~\cite{Snieder99}. Sometimes, and depending on the body of literature, they are known as process-based models or mechanistic models. In remote sensing, we refer to them as {\em radiative transfer models} as they implement the equations of energy (radiation) transfer}. This is known as the {\em forward modeling} problem. In the {\em inverse modeling} problem, the aim is to reconstruct the system state from a set of measurements (observations), see Fig.~\ref{forward_inverse}. Notationally, a forward model describing the system is generally expressed as $\x =g(\y,\bomega)$, where $\x$ is a measurement obtained by the satellite (e.g. radiance); the vector $\y$ represents the state of the biophysical variables on the Earth (which we desire to infer or predict and are often referred to as \emph{outputs} in the inverse modeling approach); $\bomega$ contains a set of controllable conditions (e.g. wavelengths, viewing direction, time, Sun position, and polarization); and $g(\cdot)$ is a function which relates $\y$ with $\x$. Such a function $g$ is typically considered to be nonlinear, smooth and continuous. Our goal is to obtain an inverse model, $f(\cdot) \approx g^{-1}(\cdot)$, parameterized by $\btheta$, which approximates the bio-geo-physical variables $\y$ given the data $\x$ received by the satellite, i.e. $\hat\y = f(\x,\btheta)$. Radiative transfer models (RTMs) are typically used to implement the forward direction~\cite{jacquemoud00,verhoef03}. However, inverting RTMs directly is very complex because the number of unknowns is generally larger than the number of independent radiometric information~\cite{Liang08}. Also, estimating physical parameters from RTMs is hampered by the presence of high levels of uncertainty and noise, primarily associated to atmospheric conditions, sensor calibration, sun angle, viewing geometry, as well as the poor sampling of the parameter space in most of the applications. All these issues translate into inverse problems where deemed similar spectra may correspond to very diverse solutions. This gives raise to undetermination and ill-posed problems.

Methods for model inversion and parameter retrieval can be roughly separated in three main families: statistical, physical and hybrid methods~\cite{CampsValls11mc}. {\em Statistical inversion} predicts a biogeophysical parameter of interest using a training dataset of input-output data pairs coming from concurrent measurements of the parameter of interest (e.g. leaf area index -LAI-) and the corresponding satellite observations (e.g. radiances or reflectances). Statistical methods typically outperform other approaches, but ground truth measurements involving a terrestrial campaign are necessary. On the contrary, {\em physical inversion} reverses RTMs by searching for similar spectra in look-up-tables (LUTs), and assigning the parameter state corresponding to the most similar observed spectrum. This requires selecting an appropriate cost function, and generating a rich, representative LUT from the RTM. The use of RTMs to generate data sets is a common practice, and especially convenient because acquisition campaigns are very costly in terms of time, money, and human resources, and usually limited in terms of parameter combinations. Finally, {\em hybrid inversion} exploits the input-output data generated by RTM simulations and train statistical regression models to invert the RTM model. Hybrid models combine the flexibility and scalability of machine learning while respecting the physics encoded in the RTMs. Currently, kernel machines in general~\cite{CampsValls09wiley}, and Bayesian non-parametric approaches such as Gaussian Process (GP) regression~\cite{Rasmussen06} in particular, are among the preferred regression models~\cite{Verrelst20121832,CampsValls16grsm}.

These GP models have been implemented in Earth observation operational chains for the derivation of biophysical variables at global scale, such as LAI through an hybrid approach. 
In addition, multitemporal LAI retrievals were derived with similar methodology at local scale taking the advantage of remote sensing data at decametric spatial resolutions and short revisit time, such as Sentinel-2 data \cite{CamposTaberner2017,busetto2017downstream}. These features allow spatial and temporal interpretation of GP estimates and their associated uncertainties at field level which can be related with remote sensing artifacts (e.g., clouds) and crop heterogeneity (e.g., crop damages).

While hybrid inversion is practical when no \emph{in situ} data is available, intuitively it makes sense to let predictions be guided by actual measurements whenever they are present. Likewise, when only very few real \emph{in situ} measurements are available, it is sensible to incorporate simulated data from RTMs to properly ground the models. This is a novel approach considered in this work, which extends the hybrid inversion by proposing a statistical method that performs nonlinear and nonparametric inversion blending both real and simulated data. The so-called joint GP (JGP) essentially learns how to trade off noise variance in the real and simulated data~\cite{Svendsen17jgp}.

A second topic covered in this work follows an alternative pathway to {\em learn} latent functions that generated the observations using GP models. We introduce a {\em latent force model} (LFM) for GP modelling~\cite{alvarez2009latent,alvarez2013linear}. The proposed LFM-GP combines the ordinary differential equations of the forward model (through smoothing kernels) and empirical data (from {\em in situ} campaigns). The LFM presented here performs multi-output structured regression, adapts to the signal characteristics, is able to cope with missing data in the time series, and provides explicit latent functions that allow system analysis and evaluation.

Finally, we deal with the important issue of {\em emulation} of RTMs, that is {\em learning} surrogate GP models to approximate costly RTMs. The proposed Automatic Gaussian Process Emulator (AGAPE) methodology combines the interpolation capabilities of Gaussian processes (GPs) with the accurate design of an acquisition function that favours sampling in low density regions and flatness of the interpolation function. AGAPE allows building compact sets to perform efficient inverse modelling while respecting the complex physical rules encoded in RTMs. 

All in all, in this paper we will illustrate the use of GPs in standard retrieval applications. In particular, we will introduce GPs to tackle problems of hybrid modeling, extending the naive application of previous works. We formalize a full framework for Earth observation with GPs. The framework incorporates different GPs models, and extend our previous works on including temporal information in GP modeling~\cite{CamposTaberner2016b,CamposTaberner2016a}, incorporating both simulated and real data~\cite{Svendsen17jgp}, advancing in the incorporation of physical rules in the modeling through the generation of kernel functions out of differential equations~\cite{alvarez2009latent,Luengo16mlsp}, multiple output GPs to assess consistency of the predictions~\cite{Luengo16mlsp}, and to learn compact look-up-tables (LUT) and emulators of RTMs using GPs in a Bayesian optimization procedure~\cite{CampsValls17agu_emula,Martino17igarss,Martino17scia}. This work improves the previous survey in~\cite{CampsValls17scia} with an improved literature review and contextualization, as well as new experimental results on the use of GPs in precision agriculture (see \S\ref{sec:2}), new results and application to transfer learning of the joint GP (see \S\ref{sec:3}), new results for the latent force model in gap-filling problems originally introduced in~\cite{Luengo16mlsp} (see \S\ref{sec:4}), as well as more details, new formulations and experiments for the automatic emulator model, which is now fully automatic and works for multioutput problems, see \S\ref{sec:5}.

The remainder of the paper is organized as follows. We first briefly introduce the standard GP for regression in \S\ref{sec:2}. Then a Joint Gaussian Process (JGP) is proposed in \S\ref{sec:3} that exploits the regularities between real and simulated data, and provides a simple framework for incorporating physical knowledge into a GP model. We introduce LFMs in \S\ref{sec:4} for vegetation monitoring across time, and then an automatic emulator based on GPs is presented in Section \S\ref{sec:5}. We conclude in \S \ref{sec:6} with some remarks and an outline of future work.

\section{Gaussian Process Models for Inverse Modeling}\label{sec:2}

GPs are state-of-the-art tools for regression and function approximation, and have been recently shown to excel in biophysical variable retrieval by following both statistical~\cite{Verrelst20121832,CampsValls16grsm} and hybrid approaches~\cite{CamposTaberner2015,CamposTaberner2016b}. 

\subsection{A brief overview of GPs}

Let us consider a set of $n$ pairs of observations or measurements, ${\mathcal D}_n:=\{\x_i,y_i\}_{i=1}^n$, where $\x\in\Real^d$ and $\y\in\Real$, which are perturbed by an additive independent noise. The input data pairs $(\X\in\Real^{n\times d},\y\in\Real^{n\times 1})$ used to fit the inverse machine learning model $f(\cdot)$ come from either {\em in situ} field campaign data (statistical approach) or simulations by means of an RTM (hybrid approach).
We assume the following model,
\begin{equation}\label{GLR}
y_i = f(\x_i) + e_i,~e_i \sim\Normal(0,\sigma_e^2),
\end{equation}
where $f(\x)$ is an unknown latent function, $\x$ $\in$ $\Real^d$, and $\sigma_e^2$ stands for the noise variance. Defining $\y$ $=$ $[y_1, \ldots ,y_n]^\intercal$ and $\mathbf{f}$ $=$ $[f(\x_1),\ldots , f(\x_n)]^\intercal$, the conditional distribution of $\y$ given $\mathbf{f}$ becomes $p(\y | \mathbf{f}) = \mathcal{N}(\mathbf{f},\sigma_e^2\I)$, where $\I$ is the $n\times n$ identity matrix. Now, in the GP approach, we assume that $\mathbf{f}$ follows a $n$-dimensional Gaussian distribution $\mathbf{f} \sim \mathcal{N}(\bm{0}, \bK)$ \cite{bishop2006pattern}.

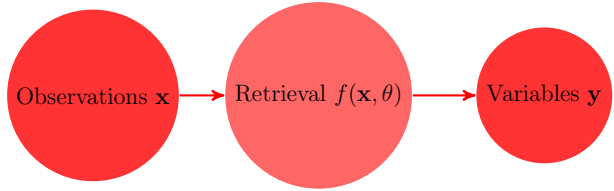
\begin{figure}[h!]
\begin{center}
\begin{tikzpicture}[->,>=stealth',scale=.75,transform shape,node distance=4cm,thick]
  \tikzstyle{every state}=[fill=red!30,draw=none,text=black]
  \node[state] (OBS)  [fill=red!80] {Observations $\x$};
  \node[state] (RET)  [right of=OBS, fill=red!60] {Retrieval $f(\x,\theta)$};
  \node[state] (VARS) [right of=RET,fill=red!80] {Variables $\y$};
  \path[solid, thick, color=red] (OBS)   edge node {} (RET);
  \path[solid, thick, color=red] (RET)   edge node {} (VARS);
\end{tikzpicture}
\end{center}
\vspace{-0.25cm}
\caption{Statistical inverse modelling. {Given a set of observations $\x$ and set of parameters $\boldsymbol{\theta}$, the statistical model $f(\x,\theta)$ provides estimations of the variables $\y$. 
In this case the model performs the inverse function of a physical model, which starting from the variables $\y$ provides the observations $\x$.}}
\label{inverse}
\end{figure}

The covariance matrix ${\bf K}$ of this distribution is determined by a kernel function with entries $\bK_{ij}=k(\x_i,\x_j)=\exp(-\|\x_i-\x_j\|^2/(2\sigma^2))$, encoding similarity between the input points \cite{Rasmussen06}. The intuition here is the following: the more similar input $i$ and $j$ are, according to some metric, the more correlated output $i$ and $j$ ought to be. Thus, the marginal distribution of $\y$ can be written as
\begin{equation}
p(\y) = \int p(\y | \mathbf{f}) p(\mathbf{f}) d \mathbf{f} = \mathcal{N}(\mathbf{0},\bm{C}_n), \nonumber
\end{equation}
where $\bm{C}=\bK + \sigma_e^2\I$. 
Now, what we are really interested in is predicting a new output $y_\ast$, given an input $\x_\ast$. The GP framework handles this by constructing a joint distribution over the training and test points,
\begin{equation}
\begin{bmatrix}
\y \\
y_\ast
\end{bmatrix}
\sim
\mathcal{N} \left( \mathbf{0},
\begin{bmatrix}
\bm{C} & \bk_\ast^\intercal \\
\bk_\ast & c_\ast
\end{bmatrix} \right),  \nonumber
\end{equation}
where $\bk_{*} = [k(\x_*,\x_1), \ldots, k(\x_*,\x_n)]^\intercal$ is an $n\times 1$ vector and $c_\ast = k(\x_*,\x_\ast) + \sigma_e^2$. Then, using standard GP manipulations, we can find the distribution over $y_\ast$ conditioned on the training data, which is a normal distribution with predictive mean and variance given by
 \begin{align}\label{eq:gppred}
  \begin{aligned}
   \mu_{\text{GP}} (\x_\ast) &= \bk_{*}^\intercal (\bK + \sigma_e^2\I_n)^{-1}\y, \\
   \sigma^2_{\text{GP}} (\x_\ast) &= c_\ast - \bk_{*}^\intercal (\bK + \sigma_e^2\I_n)^{-1} \bk_{*}.
  \end{aligned}
 \end{align}
Thus, GPs yield not only predictions $\mu_{\text{GP}\ast}$ for test
data, but also the so-called ``error-bars'', $\sigma_{\text{GP}\ast}$, assessing the uncertainty of the mean prediction. The hyperparameters $\bm{\theta}=[\sigma, \sigma_e]$ to be tuned in the GP determine the width of the squared exponential kernel function and the noise on the observations. This can be done by marginal likelihood maximization or simple grid search, attempting to minimize the squared prediction errors. In the next section we describe some practical cases regarding the use of GPs both $\mu_{\text{GP}\ast}$ and $\sigma_{\text{GP}\ast}$ in EO.

\subsection{GPs for model inversion in precision agriculture}\label{sectionPA}

In this section, we show examples of the GP regression (GPR) model utility in real world applications related to precision farming/agriculture from remote sensing data. In this case, GPR was used for inverting the PROSAIL radiative transfer model (thus following a hybrid approach). PROSAIL simulates leaf reflectance for the optical spectrum, from 400 to 2500 nm with a 1 nm spectral resolution, as a function of biochemistry and structure of the canopy, its leaves, the background soil reflectance and the system geometry. The leaf and canopy variables as well as the soil brightness parameter, were generated following a PROSAIL site-specific parameterization to constrain the model to Mediterranean rice areas~\cite{CamposTaberner2016b}. Firstly, PROSAIL was run in forward mode in order to build a database composed of pairs of simulated Sentinel-2 spectra and associated LAI values. A total number of $2000$ simulations were computed in such a way the obtained spectra and LAI values covered the expected season of rice crops as well as their management (agricultural practice). Then, the database (often called look-up table) was used for training the GPR model, which was then used for estimating LAI using real Sentinel-2 imagery. Hence, every time a Sentinel-2 image was available, the corresponding LAI map was derived. This procedure was conducted between mid-May until early-October thus completely covering the rice season. As result we derived 11 Sentinel-2 LAI maps.

\begin{figure}[t!]
\centering
\includegraphics[width=10cm]{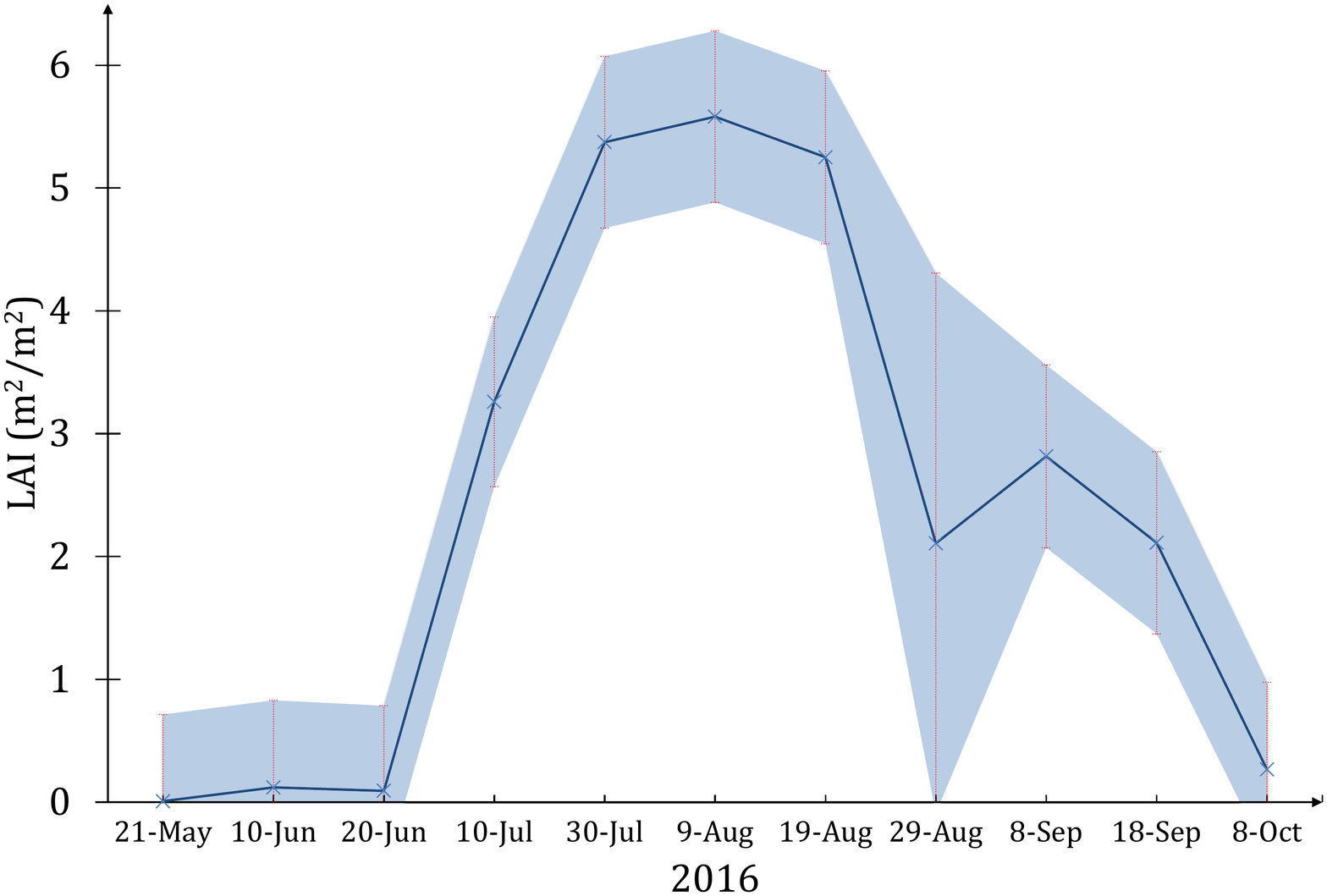}
	\vspace{-0.2cm}
	\caption{Temporal evolution of GPR LAI estimates and associated uncertainty (shaded space) over a rice pixel.}
	\label{fig:cloud1}
\end{figure}

\begin{figure}[t!]
    \centering
    \includegraphics[width=10cm]{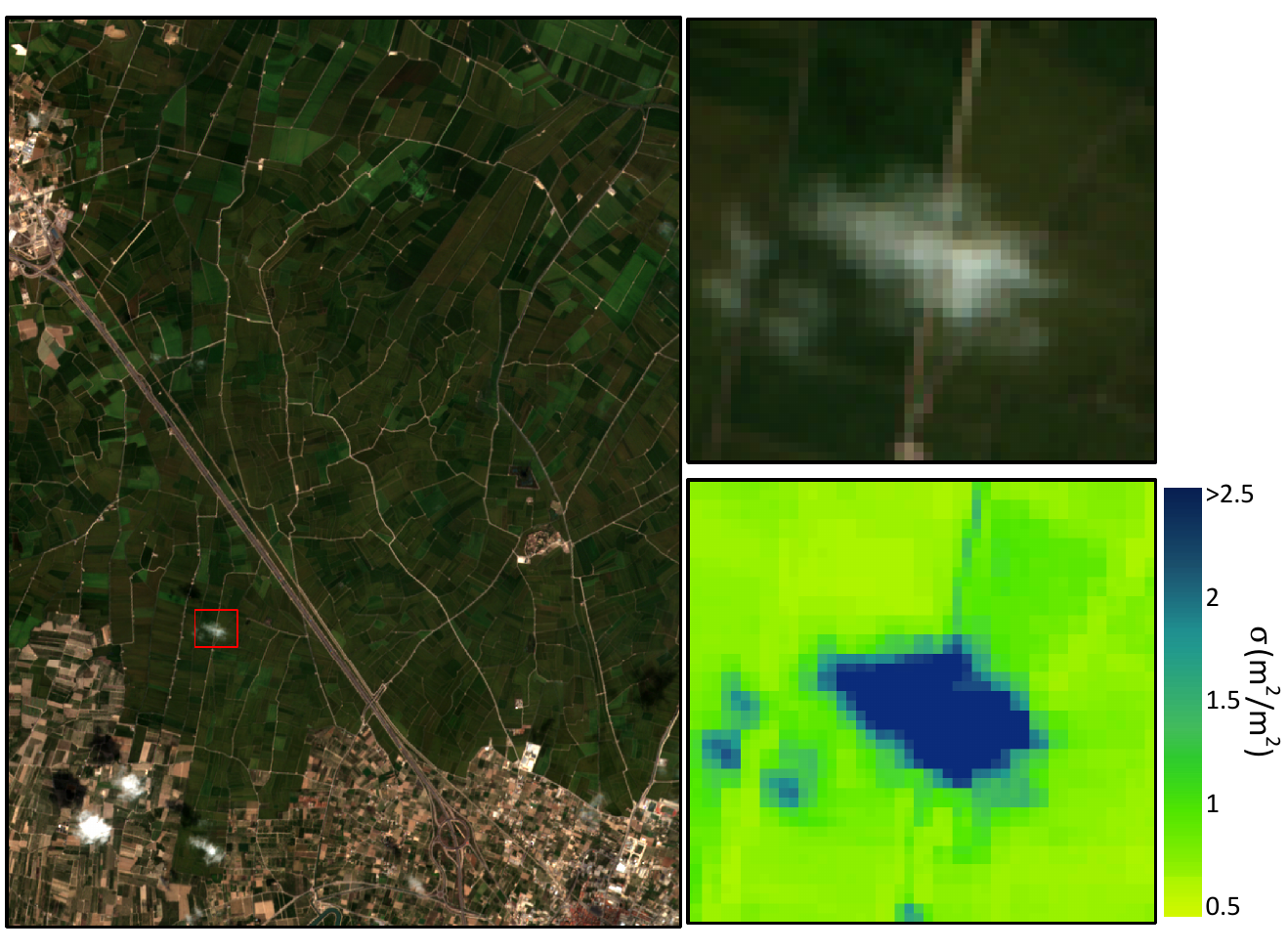}			  	\vspace{-0.2cm}
	\caption{Sentinel-2 RGB composite over rice fields near Valencia (Spain) on August, 29th 2016 and the corresponding LAI uncertainty over an undetected cloud.}
	\label{fig:cloud2}
\end{figure}

Figure \ref{fig:cloud1} shows the LAI evolution over a rice field which is in accordance with rice plant evolution. It is worth mentioning that an unexpected drop was detected on August 29th. LAI decreased too much on this date: a LAI decrease about 3 in a 10-day period does not correctly characterize the typical rice LAI behaviour. Moreover, if we observe the temporal evolution of the GP predictive variance, $\sigma_\text{GP}$, values remain virtually constant at a value about 0.8. However, $\sigma_\text{GP}$ dramatically increases (up to 3) on the date that the unexpected drop was observed.  
Figure~\ref{fig:cloud2} provides a Sentinel-2  
a map for the predictive variance on August 29th where undetected clouds presented very high values as compared with cloud free rice fields. The lower confidence (higher prediction uncertainty) is associated with spectra non-represented in the PROSAIL training database. Therefore, non-vegetated surfaces, such as clouds, present higher prediction uncertainty (lower confidence). This assessment of $\sigma_\text{GP}$ can be useful to properly weight estimates with low confidence when used by crop modelers, and also to to improve cloud masks. 

\begin{figure}[t!]
\centering
\includegraphics[width=12cm]{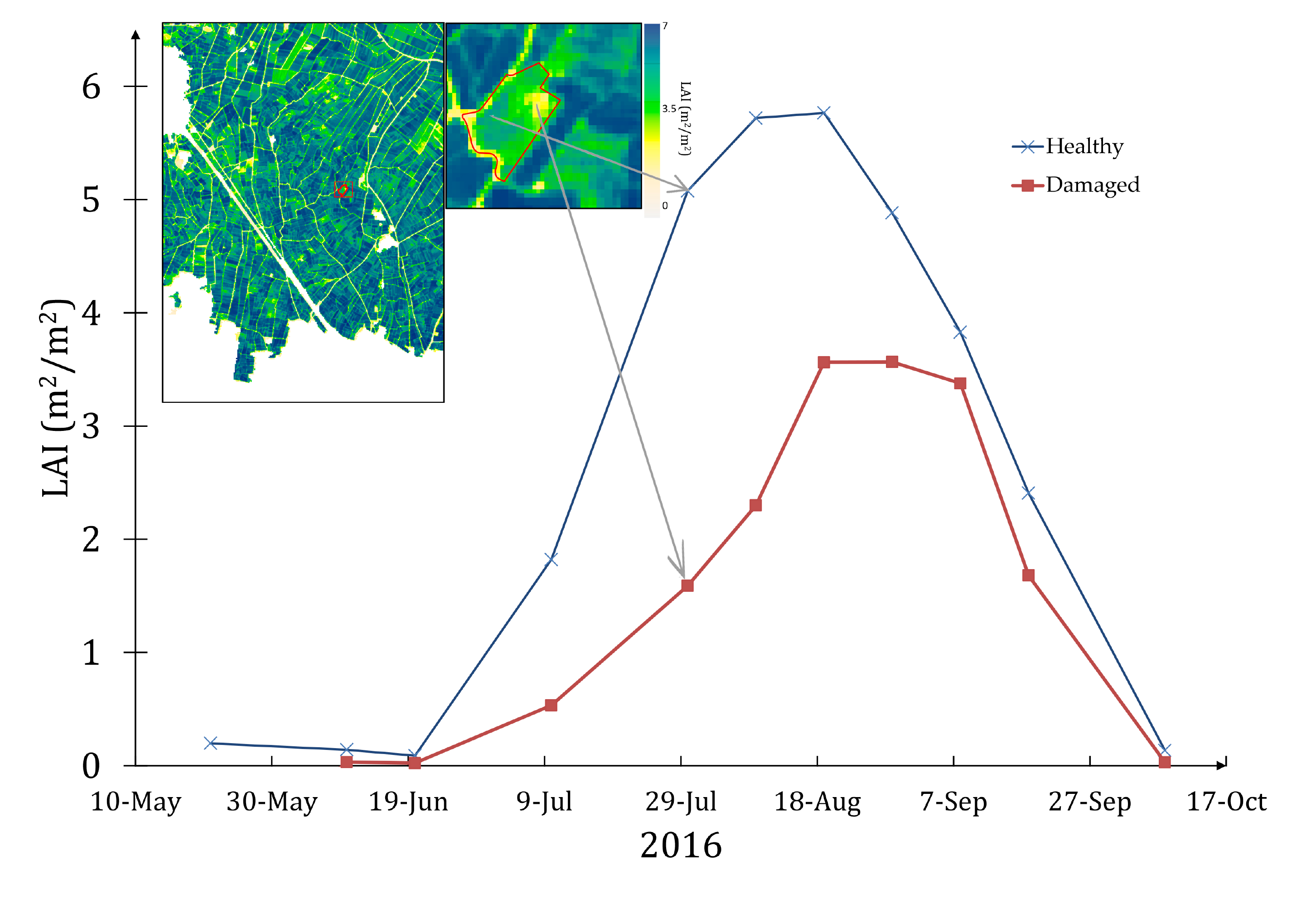}
	\vspace{-0.2cm}
	\caption{Temporal evolution of GPR LAI estimates and Sentinel-2 LAI map on July, 29 2016. Red line corresponds to a damaged pixel while the blue one corresponds to a healthy pixel within the same field.}
	\label{fig:damages}
\end{figure}

The spatio-temporal detail of the derived maps due to the use of Sentinel-2 data (10 m spatial and 10-day temporal resolution), allows intra-field and multi-temporal analysis useful for crop assessment \cite{CamposTaberner2017}. These features allow the identification of significant different values within the same rice field. Intra-field LAI differences are mainly due to the heterogeneity of the field related with non-homogeneous agro-practices. The retrieved high-resolution LAI estimates can be used to continuously monitor the cropping season and to detect crop growth anomaly linked with potential crop damage. In particular, Fig. \ref{fig:damages} exhibits the temporal evolution of two pixels within the same rice field. The blue line corresponds to the LAI evolution of a healthy pixel and the red one describes the temporal behaviour of a pixel located in the same rice field but affected by a rice disease. According to the temporal profiles, the anomalous LAI behaviour started on the beginning of the season impacting in the LAI values mainly in the rice development stage. This information was corroborated by \emph{in situ} observations. Overall, this kind of analysis and assessment can be used to early derive anomalies maps related with crop damages which could be used by farmers in order to apply agro-practices for mitigating yield loss. 

\section{Forward and Inverse Joint GP Models}\label{sec:3}

\subsection{Notation and formulation}

Let us now assume that the previous dataset $\dataset_n$ is formed by two disjoint sets: one set of $r$ real data pairs, $\dataset_r=\{(\x_i,y_i)\}_{i=1}^r$, and one set of $s$ RTM-simulated pairs $\dataset_s=\{(y_j,\x_j)\}_{j=r+1}^n$, so that $n=r+s$ and $\dataset_n=\dataset_r\cup \dataset_s$. In matrix form, we have $\X_r\in\Real^{r\times d}$, $\y_r\in \Real^{r\times 1}$, $\X_s\in\Real^{s\times d}$ and $\y_s\in \Real^{s\times 1}$,  containing all the inputs and outputs of $\dataset_r$ and $\dataset_s$, respectively. Finally, the $n\times 1$ vector $\y$ contains all the $n$ outputs, sorted with the real data first, followed by the simulated data.
Now, we define a different model where the observation noise depends on the origin of the data: $\sigma_e^2$ for real observations ($\x_i \in \dataset_r$) or $\sigma_e^2/\gamma$ for RTM simulations ($\x_i \in \dataset_s$),
where the parameter $\gamma>0$ accounts for the importance of the two sources of information relative to each other. 

The resulting distribution of $\y$ given $\mathbf{f}$ is only slightly different from that of the regular GP, namely $p(\y|\mathbf{f}) = \mathcal{N}(\mathbf{f}, \sigma_e^2 \mathbf{V})$ where $\mathbf{V}$ is an $n\times n$ diagonal matrix in which the first $r$ diagonal elements are equal to $1$ and the remaining $s$ are equal to $\gamma^{-1}$: $\mathbf{V}=\text{diag}(1,\ldots,1,\gamma^{-1},\ldots, \gamma^{-1})$. 
The predictive mean and variance of a test output $y_\ast$, conditioned on the training data, then becomes

 \begin{align} \label{eq:pred}
  \begin{aligned}
   \mu_{\text{JGP}} (\x_\ast) &= \bk_{*}^\intercal (\bK + \sigma_e^2\mathbf{V})^{-1}\y, \\
   \sigma^2_{\text{JGP}} (\x_\ast) &= c_\ast - \bk_{*}^\intercal (\bK + \sigma_e^2\mathbf{V})^{-1} \bk_{*}.
  \end{aligned}
 \end{align}
 
\noindent Note that when $\gamma=1$ the standard GP formulation is obtained. Otherwise $\gamma$ acts as an extra regularization term accounting for the relative importance of the real and the simulated data points. Selection of the hyperparameters of the JGP, $\btheta = [\sigma,\sigma_n,\gamma],$ is central to the effectiveness of the model, since what we are really interested is in performing predictions on the real data. We therefore maximize the \textit{pseudo}-likelihood \cite{Rasmussen06} of the \textit{real} data only:

\begin{equation}\label{LOO1}
	L_{\mbox{\footnotesize }} (\X,\y, \boldsymbol{\theta}) = \sum_{i=1}^{r} \log p(y_i | \X_{\backslash i},\y_{\backslash i}, \boldsymbol{\theta}),
\end{equation}

\noindent where we sum over the log-likelihood of each \textit{real} data point given the remaining training data.
{The sub-index $\backslash i$ represents the remaining training data}.
The log-likelihood of a single point $i$ given the remaining data is

\begin{align*}
\log p(y_i | \X_{\backslash i},\y_{\backslash i}, \boldsymbol{\theta} ) =
-\frac{1}{2}\log 2\pi \sigma^2_i - \frac{(y_i - \mu_i )^2}{\sigma^2_i},
\end{align*}

\noindent where $\mu_i$ and $\sigma^2_i$ are computed using \eqref{eq:pred} {with all $r+s$ datapoints except the $i$'th}. By optimizing hyperparameters in this way, $\gamma$ becomes a measure of how useful the simulated data is in predicting the real data.

\subsection{Experimental Results} 
\label{sec:expres1}

We are concerned about the prediction of leaf area index (LAI) parameter from space, a parameter that characterizes plant canopies and is roughly defined as the total needle surface area per unit ground area. Non-destructive real LAI data were acquired over Elementary Sampling Units (ESUs) within rice fields in Spain, Italy and Greece during field campaigns in 2015 and 2016, i.e. 6 datasets. The temporal frequency of the campaigns was approximately 10 days starting from the very beginning of rice emergence (early-June) up to the maximum rice green LAI development (mid-August). LAI measurements were acquired using a dedicated smartphone app (PocketLAI), which uses both the smartphone's accelerometer and camera to acquire images at 57.5$^\circ$ below the canopy and computes LAI through an internal segmentation algorithm~\cite{CamposTaberner2015}. The center of the ESU was geo-located for later matching and association of the mean LAI estimate with the corresponding satellite spectra. We used Landsat 8 surface reflectance data over each area corresponding to the dates of measurements' acquisition. The resulting datasets contain a number of {\em in situ} measurements in the range of $70$-$300$ depending on the country and year. On the other hand, three simulated datasets of $s=2000$ pairs of Landsat 8 spectra and LAI, with characteristics corresponding to the relevant rice area, were obtained running the PROSAIL RTM in forward mode following the similar procedure described in Section \ref{sectionPA}, but in this case we simulated Landsat-8 spectra instead of {Sentinel-2}.

Two types of experiments were conducted. In the first one, we investigate, for each of the 6 datasets, how including simulated data might improve predictions in a regular $10$-fold cross-validation scheme. 
In the second experiment, we explore how simulated data might help prediction of LAI in one site, given that one only has access to data from a different site. This is a quite habitual situation, often referred to as {\em domain adaptation} or {\em transfer learning} in machine learning~\cite{CampsValls11mc,Tuia16plosone}; or simply as {\em model transferability} in remote sensing applications. In this case, we use datasets from 2016.
We shall refer to these experiment types as \textit{same-site} and \textit{cross-site} respectively.

\subsubsection{Same-site experiments}
\begin{figure}[t!]
	\small
	\begin{center}
			\hspace{-0.1cm}\includegraphics[width=1.0\textwidth]{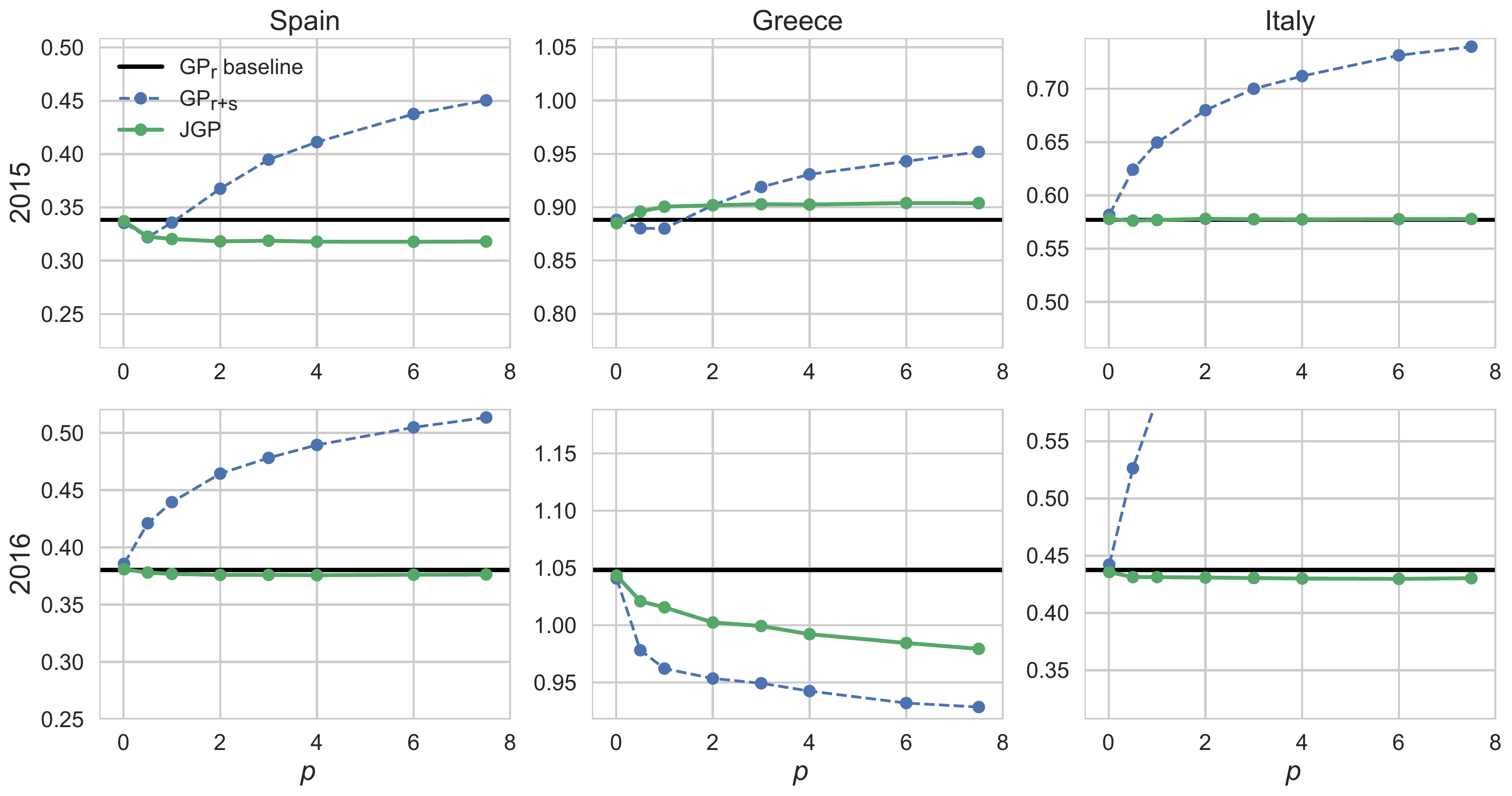} \\
	\end{center}
	\vspace{-0.5cm}
	\caption{Performance comparison (RMSE) for different ways of including simulated data.
	{$p=s/r$ is the ratio between simulated ($s$) and real ($r$) data}.
	The JGP and the regular GP, trained on a dataset of real and simulated data pooled together (i.e. the GP$_{r+s}$), are compared to the base line of the GP trained exclusively on real data. RMSE is shown for the different sites, campaign dates and simulated-to-real data ratios. As the scale is constant over the plots for better comparison, it was omitted how the GP$_{r+s}$ RMSE monotonically increases and reaches 0.85 for $p=8$ from the plot in Italy 2016.}
	\label{fig:prob1}
\end{figure}

We assessed the performance of the JGP for different amounts of simulated data. We compare to a regular GP model (see Sec. \ref{sec:2}) which only has access to real data, which we will refer to as GP$_{\text{r}}$, and a different regular GP model which has access to a training set of both simulated and real data. This, more naive approach of including simulated data, which does not distinguish between data sources, is referred to as GP$_{\text{r+s}}$.

{Figure \ref{fig:prob1} shows the effect of the ratio between simulated and real data points, $p=s/r$, on the RMSE evaluated using 10-fold crossvalidation.} The JGP behaves in different ways on different datasets. There are cases where a $\gamma$ value close to 0 is fitted, the simulated data is largely ignored, and it follows the GP$_{\text{r}}$ baseline. For the datasets where this does not happen, we see that a $p\sim 1$ is enough to produce an effect. This is worth noting as the inversion of the kernel matrix, needed to train the JGP, scales in time complexity with the number of samples cubed, $\mathcal{O}(n^3) = \mathcal{O}((r+s)^3)$.

In the case of Greece 2015, an average increase in RMSE is observed which, percentage-wise is around $\sim$ 1 \%. In Spain 2015 and Greece 2016, a decrease in RMSE of around $\sim$ 5 \% can be observed. Interestingly, we see that the naive inclusion of simulated data (the GP$_{\text{r+s}}$ scheme) generally leads to an increase in error, except for the case of Greece 2016. This hints towards the fact that the GP$_{\text{r+s}}$ can perform better than the JGP approach when simulated data is of high quality, as it is less conservative.
Overall, the JGP appears to be a safe way to include simulated data, at worst increasing RMSE by $\sim$ 1 \% in one dataset, and at best decreasing it by $\sim$ 5 \%. This is made possible by the hyperparameter fitting procedure which attempts to assess whether the simulated data is useful or confusing for prediction.

\if
The gain in accuracy was measured as the reduction in root mean square error (RMSE gain [\%] = $100\times$(RMSE$_{\text{GP}}$-RMSE$_{\text{JGP}}$)/RMSE$_{\text{GP}}$). We evaluated performance in the 6 datasets generated for different countries (SP, GR, IT) and years (2015, 2016). Figure \ref{fig:prob1} shows the effect of the ratio between simulated and real data points $p = s / r$ on the RMSE gain evaluated using 10-fold crossvalidation. When no simulated data is used, the JGP model reduces to the standard GP model, but when introducing an amount of PROSAIL-datapoints similar to the amount of real datapoints, i.e. $p\sim 1$, a noticeable gain is for datasets gathered in 2016. In the case of the data from Spain, the gain appears rather stable (between 6 and 2\% in 2015 and 2016 respectively) after reaching a ratio of $p=2$, indicating what size of the simulated dataset is needed for an increase in accuracy. The results for Greece and Italy, however, show that the use of simulated data attempting to fill in the under-represented domain of the real data, is not always useful.
\fi

\subsubsection{Cross-site experiments}

\begin{figure}[t!]
\centering
\includegraphics[width=\textwidth]{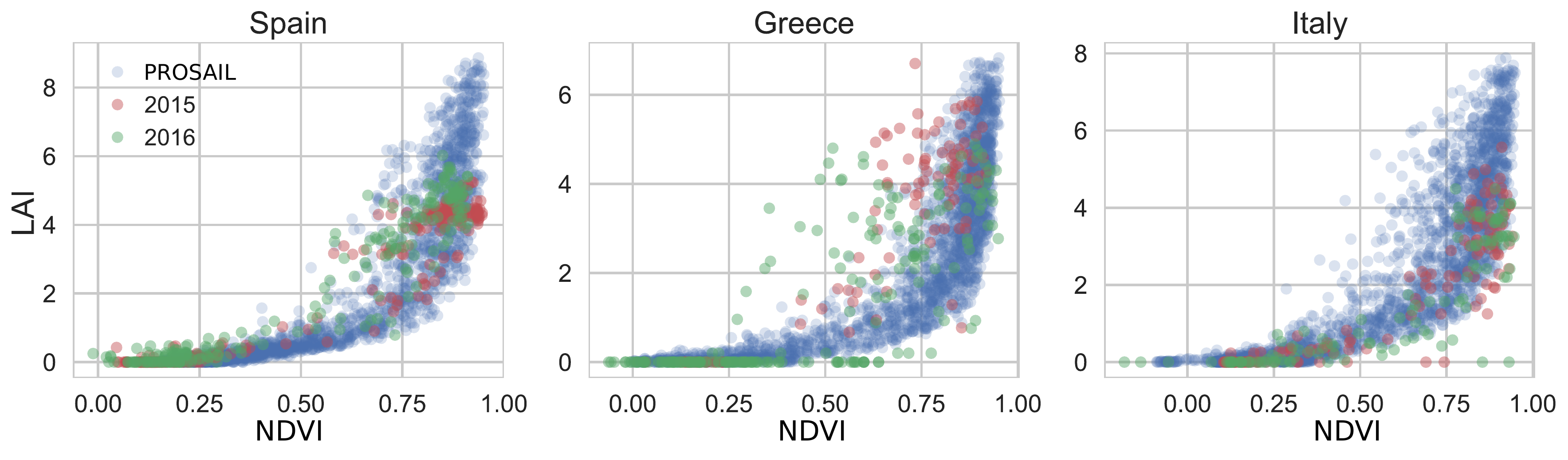}
\caption{Scatterplots in the NDVI-LAI representation space of the real and RTM-simulated data for all sites and acquisition campaigns (2015, 2016).}
\label{fig:ndviscatter}
\end{figure}

We turn now to the question of whether a regression model, trained on data from one particular site, can be useful for prediction at a different one. The experiment is not incidental, but of practical concern and implications, as it is expensive to collect data, limiting the amount of real data available. It comes down to a question of how similar the distributions over the output variable LAI, given the input variables, are across sites.

In order to visualize the 6 datasets we plot them in the NDVI-LAI representation space in Fig. \ref{fig:ndviscatter}, along with their simulated counterparts. The PROSAIL simulated data exhibits the well known exponential relation between NDVI and LAI, and a similar trend is visible in the real datasets.
The data distribution that stands the most out, both with respect to simulated and other real datasets, is that pertaining to the Greek site. 
This is, as we shall see, what determines how well a model trained on that data will perform on other datasets.

\begin{figure}[t!]
\centering
\includegraphics[width=\textwidth]{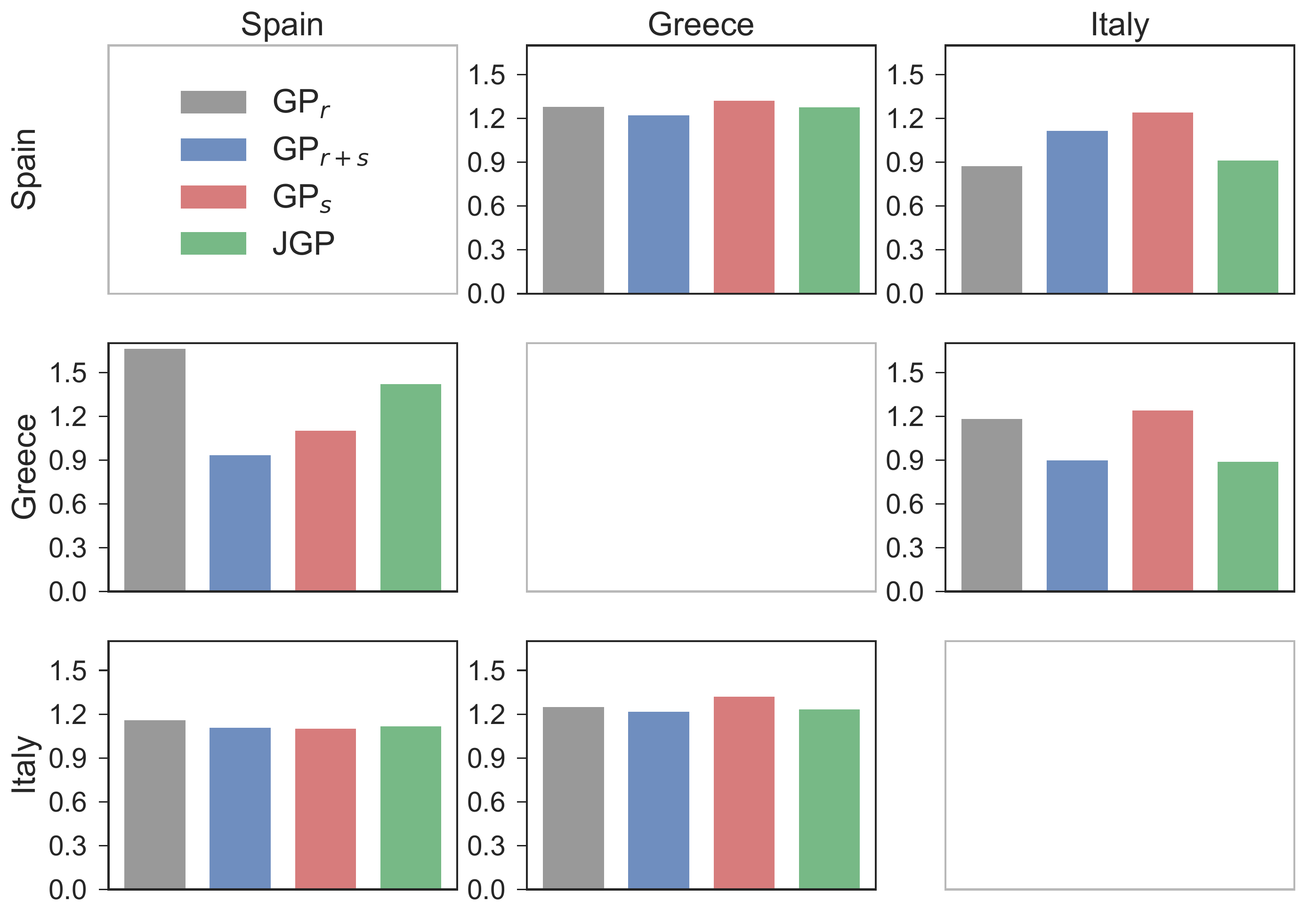}
\caption{Performance (RMSE) of the different approaches to cross-site learning, where rows and columns indicate the source and target datasets respectively.}
\label{fig:transfer}
\end{figure}

We consider the following strategies for predicting on a \textit{target} site, having only data from a \textit{source} site. One might train a GP with the available \textit{real} source data, denoted GP$_{\text{r}}$ in this section. 
Otherwise, we might have some knowledge about the target site that we can use to create an RTM-simulated database. This strategy of training only on simulated data will be referred to as GP$_{\text{s}}$ here. 
Finally, one can try to combine the real data from the source site, with simulated target site data. This could be attempted through training a normal GP on the union of these two datasets, i.e. a  GP$_{\text{r+s}}$ model, or through the JGP.

Figure \ref{fig:transfer} shows how these methods compare, where row names indicate the train/source site and the column denotes the test/target site. Thus, the RMSE of the GP$_{\text{s}}$ is constant across columns. We see how the fact that the simulated data distribution poorly matches that of the real data in Greece and Italy (see Fig. \ref{fig:ndviscatter}) is reflected in the RMSE for the GP$_{\text{s}}$ approach in the two rightmost columns of Fig. \ref{fig:transfer}. 
Conversely, we see from the second row how inclusion of simulated data in some form is very useful for predicting in other sites when having access only to the real dataset from Greece. We note here, about the JGP, that it is the only method that consistently performs better than the simple GP$_{\text{r}}$ strategy. 
In conclusion, the JGP can be said to be a safe approach to include simulated data for non-linear regression.

\section{Inverse Modelling with Latent Force Models}\label{sec:4}
 
\begin{figure}[t!] 
\begin{center}
\begin{tikzpicture}[->,>=stealth',scale=.75,transform shape,node distance=3cm,thick]
  \tikzstyle{every state}=[fill=red!30,draw=none,text=black]
  \node[state] (OBS)  [fill=red!80] {Observations $\x$};
  \node[state] (LF2)  [right of=OBS,fill=red!60] {LF $f_2(\x)$};
  \node[state] (LF1)  [above of=LF2,fill=red!60] {LF $f_1(\x)$};
  \node[state] (LFR)  [below of=LF2,fill=red!60] {LF $f_R(\x)$};
  \node[state] (RET)  [right of=LF2, fill=red!60] {Retrieval $f(\x,\theta)$};
  \node[state] (VARS) [right of=RET,fill=red!80] {Variables $\y_q$};
  \path[solid, bend left=25, thick, color=red] (OBS) edge node {} (LF1);
  \path[solid, thick, color=red] (OBS) edge node {} (LF2);
  \path[solid, bend right=25, thick, color=red] (OBS) edge node {} (LFR);
  \path[solid, bend left=25, thick, color=red] (LF1) edge node {} (RET);
  \path[solid, thick, color=red] (LF2) edge node {} (RET);
  \path[solid, bend right=25, thick, color=red] (LFR) edge node[xshift=-1.8cm,yshift=0.8cm] {\LARGE$\vdots$} (RET);
  \path[solid, thick, color=red] (LF2)   edge node {} (RET);
  \path[solid, thick, color=red] (RET)   edge node {} (VARS);
\end{tikzpicture}
\end{center}
\vspace{-0.5cm}
\caption{Inverse modeling with latent forces. {In this case, the statistical inversion model does not depend directly on the inputs, but on a set of {\em a priori} unknown independent latent functions that describe the underlying physical model.}}
\label{lfm}
\end{figure}
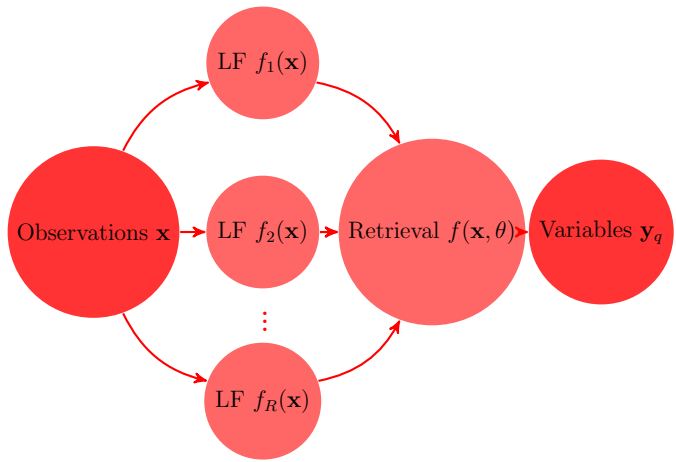

In this second case study, we are interested in inverse modelling from real {\em in situ} data, {\em learning} not only an accurate retrieval model but also the physical mechanism that generated the input-output observed relations without even accessing any RTM, see Fig.~\ref{lfm}. Here, we assume that our observations correspond simply to the temporal variable, $\x\sim t$, so the latent functions are defined in the time domain, $f_r(t)$. Nevertheless, extension to multidimensional objects such as radiances is straightforward by using different kernels. Notationally, let us consider a multi-output scenario with $Q$ correlated observed time series, $y_q(t)$ for $1 \le q \le Q$, and let us assume that we have $n$ samples available for each of these signals, taken at sampling points $t_i$, s.t. $y_q[i] = y_q(t_i)$ for $1 \le i \le n$.
This is the \emph{training set}, which is composed of an input vector, $\vec{t}=[t_1,\ \ldots,\ t_n]^{\intercal}$, and an output matrix, $\vec{Y}=[\vec{y}_1,\ \ldots,\ \vec{y}_Q]$ with $\vec{y}_q=[y_q[1],\ \ldots,\ y_q[n]]^{\intercal}$. We aim to build a GP model for the $Q$ outputs that can be used to perform inference on the \emph{test set}: $\widetilde{\vec{t}}=[\widetilde{t}_1,\ \ldots,\ \widetilde{t}_m]^{\intercal}$ and $\widetilde{\vec{Y}}=[\widetilde{\vec{y}}_1,\ \ldots,\ \widetilde{\vec{y}}_Q]$ with $\widetilde{\vec{y}}_q=[\widetilde{y}_q[1],\ \ldots,\ \widetilde{y}_q[m]]^{\intercal}$ and $\widetilde{y}_q[m'] = y_q(\widetilde{t}_{m'})$ for test inputs at $t_{m'}$.

\subsection{Formulation} 

\begin{figure}[!htb]
\includegraphics[width=0.9\textwidth]{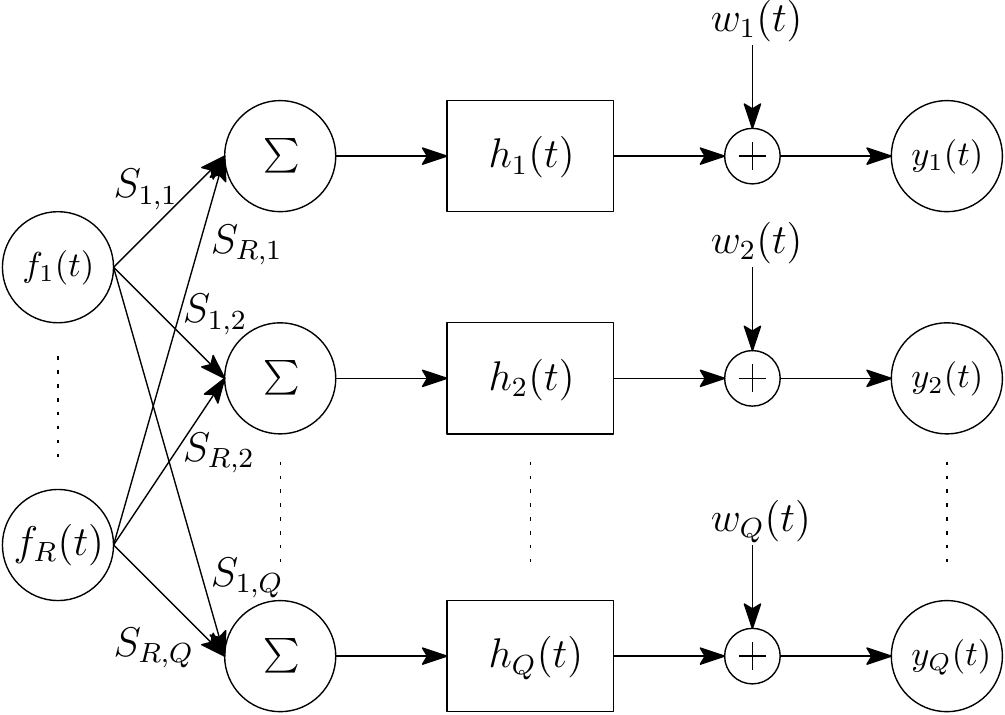}
\caption{GP-LFM relating inputs (latent forces) and outputs (observations).}
\label{lfm2}
\end{figure}

Let us assume that a set of $R$ independent latent functions (LFs), $f_r(t)$ with $1 \le r \le R$, are responsible for the observed correlation between the outputs. Then, the cross-correlation between the outputs arises naturally as a result of the coupling between the set of independent LFs, instead of being imposed directly on the set of outputs. Let us define the form of these latent functions and the coupling mechanism between them. In this work, we model the LFs as zero-mean Gaussian processes (GPs), and the coupling system emerges through a linear convolution operator described by an \emph{impulse response}, $h_{q}(t)$, as follows:
\begin{align}
	y_{r,q}(t) & = \linOpSpecific{q}{t}{f_r(t)} = f_r(t) * h_q(t) = \int_{0}^{t}{f_r(\tau) h_q(t-\tau) \textrm{d}\tau},
\label{eq:pseudoOutputsSpecific}
\end{align}
where $\linOpSpecific{q}{t}{f_r(t)}$ indicates the linear operator associated to the linear convolution of the latent force $f_r(t)$ with the \emph{smoothing kernel} $h_q(t)$.
As shown in Fig.~\ref{lfm2}, the outputs are finally obtained as a linear weighted combination of these pseudo-outputs plus an additive white Gaussian noise (AWGN) term:
\begin{equation}
	y_q(t) = \sum_{r=1}^{R}{S_{r,q} y_{r,q}(t)} + w_q(t),
\label{eq:outputsSpecific}
\end{equation}
where $S_{r,q}$ represents the coupling strength between the $r$-th LF and the $q$-th output, and $w_q(t) \sim \mathcal{N}(0,\eta_q^2)$ is the AWGN term. In practice, we consider only the squared exponential auto-covariance function for the LFs, $k_{f_r f_r}(t'-t) \propto \exp(-\frac{(t'-t)^2}{2 \ell_r^2})$, where the hyperparameter $\ell_r$ controls the length-scale of the process.

\noindent The smoothing kernel encodes our knowledge about the linear system (that relates the unobserved LFs and the outputs), and can be based on basic physical principles of the system at hand (as in \cite{alvarez2009latent,alvarez2013linear}) or selected arbitrarily (as in \cite{higdon2002space,boyle2004dependent}). In this paper, we consider the Gaussian smoothing kernel, $h_q(t) \propto \exp(-\frac{t^2}{2 \nu_q^2})$. Since the LFs are zero-mean GPs, the noise is zero-mean and Gaussian, and all the operators involved are linear, the joint LFs-output process is also a GP.
Therefore, the mean function of the $q$-th output is $\mu_{y_q}(t) = 0$, whereas the cross-covariance function between two outputs is
\begin{align}
	k_{y_p y_q}(t,t') = \sum_{r=1}^{R}{S_{r,p} S_{r,q} \linOpSpecific{p}{t}{\linOpSpecific{q}{t'}{k_{f_r f_r}(t,t')}}} + \eta_q^2 \delta[p-q] \delta[t'-t],
\label{eq:crossCovarianceOutputs}
\end{align}
where the term $\linOpSpecific{p}{t}{\linOpSpecific{q}{t'}{k_{f_r f_r}(t,t')}}$ denotes the application of the convolutional operator twice to the autocorrelation function of the LFs, which results in the following double integral:
$$
\linOpSpecific{p}{t}{\linOpSpecific{q}{t'}{k_{f_r f_r}(t,t')}} = \int_{0}^{t} \int_{0}^{t'} h_p(t-\tau) h_q(t'-\tau') k_{f_r f_r}(\tau,\tau') \diff \tau' \diff \tau.
$$
Finally, the cross-correlation between the LFs and the outputs readily gives $k_{f_r y_q}(t,t') = S_{r,q} \linOpSpecific{q}{t'}{k_{f_r f_r}(t,t')}$, which involves a single one-dimensional integral.
All integrals can be solved analytically when both the LFs and the smoothing kernel have a Gaussian shape.

Learning hyperparameters through marginal log-likelihood maximization is very challenging because of its complicated dependence on the hyperparameters $\thetavec=[\nu_q,l_r,\sigma,\sigma_n,\eta_q]$. We propose to solve the problem through a stochastic gradient descent technique, the scaled conjugate gradient~\cite{nabney2002netlab}. Once the hyperparameters $\thetavec$ of the model have been learned, inference proceeds by applying standard GP regression formulas~\cite{Rasmussen06} (cf. \S\ref{sec:2}). Now, since the conditional PDF is Gaussian, the minimum mean squared error (MMSE) prediction is simply given by the conditional mean:
\begin{equation}
	\hat{\vec{y}} = \muvec_{\tilde{\vec{y}} \vert \vec{y}}
		= \vec{K}_{\tilde{\vec{y}} \vec{y}} \vec{K}_{\vec{yy}}^{-1} \vec{y},
\label{eq:mapPredictor}
\end{equation}
where $\hat{\vec{y}} = [\hat{\vec{y}}_1^{\intercal},\ \ldots,\ \hat{\vec{y}}_Q^{\intercal}]$ is the vectorized version of the inferred outputs, which can be expressed in matrix form as $\hat{\vec{Y}}=[\hat{\vec{y}}_1,\ \ldots,\ \hat{\vec{y}}_Q]$ with $\hat{\vec{y}}_q=[\hat{y}_q[1],\ \ldots,\ \hat{y}_q[m]]^{\intercal}$ and $\hat{y}_q[m'] = \hat{y}_q(\widetilde{t}_m')$.

\subsection{Experimental Results}

We are concerned about multiple time series of two (related) biophysical parameters, LAI and fraction of Absorbed Photosynthetically Active Radiation (fAPAR), in the locations of the experiments described in Section~\S\ref{sec:expres1}. We focus on a set of representative rice pixels of each area, thus allowing us to observe the inter-annual variability of rice from 2003 to 2013 at a coarse spatial resolution (2 Km), which is useful for regional vegetation modelling.

\subsubsection{Learning the LAI-fAPAR relationships}

In this section, we explore the LAI vs. fAPAR relationship, which is usually modeled using the following exponential model, as largely observed in the literature~\cite{MYNENI2002214}: 
\begin{equation}
	\fapar = 1 - \exp(\alpha \times \lai).
\label{eq:fapar_vs_lai}
\end{equation}
In order to determine whether the GP-LFM is able to capture this well-known relationship, we train the model using the multi-output time series composed of all the available LAI and fAPAR data from the MODIS sensor for Spain and Italy from the beginning of 2003 until the end of 2013 (i.e., $N=506$ and the number of outputs is $Q=4$).
After removing truly missing data (marked with negative values in the original time series) this results in 2006 training samples (506 samples for each of the time series from Spain and 497 samples for the Italian ones).
We have experimented with a variable number of latent forces, $R \in \{1,2,3\}$.
Table \ref{tab:mse_full} shows the quantitative results in terms of mean squared error (MSE),
\begin{equation}
	\mse_q = \frac{1}{N_q}\sum_{n=0}^{N_q-1}{(y_q[n]-\hat{y}_q[n])^2},
\label{eq:mse}
\end{equation}
and normalised MSE,
\begin{equation}
	\rmse_q(\%) = \frac{\mse_q}{\frac{1}{N_q}\sum_{n=0}^{N_q-1}{y_q^2[n]}} \times 100,
\label{eq:rmse}
\end{equation}
where $y_q[n]$ denotes the true value of the $n$-th sample from the $q$-th time series, $\hat{y}_q[n]$ is the value predicted by the model, $N_q \le N$ is the number of samples available for the $q$-th time series ($N_1 = N_2 = 506$ and $N_3 = N_4 = 497$, as mentioned above) and $q=1,\ldots,Q=4$.

\begin{table}[!htb]
	\setlength{\tabcolsep}{2pt}
	\centering
	\begin{tabular}{|c|c|c||c|c|}
		\hline
		\multicolumn{5}{|c|}{MSE (NMSE)} \\
		\hline
		$R$ & LAI (ES) & LAI (IT) & fAPAR (ES) & fAPAR (IT) \\
		\hline
		1 & 0.1139 (2.08 \%) & 0.2422 (5.97 \%) & 0.0080 (4.02 \%) & 0.0046 (2.49 \%) \\
		2 & 0.0548 (1.00 \%) & 0.1636 (4.03 \%) & 0.0013 (0.67 \%) & 0.0039 (2.11 \%) \\
		3 & 0.0012 (0.02 \%) & 0.1657 (4.09 \%) & 0.0002 (0.10 \%) & 0.0025 (1.38 \%) \\
		\hline
	\end{tabular}
	\caption{Absolute and normalised MSE using the full dataset for $R \in \{1,2,3\}$ LFs.}
	\label{tab:mse_full}
\end{table}

\begin{figure*}[!t]
	\centering
	\includegraphics[width=0.32\textwidth]{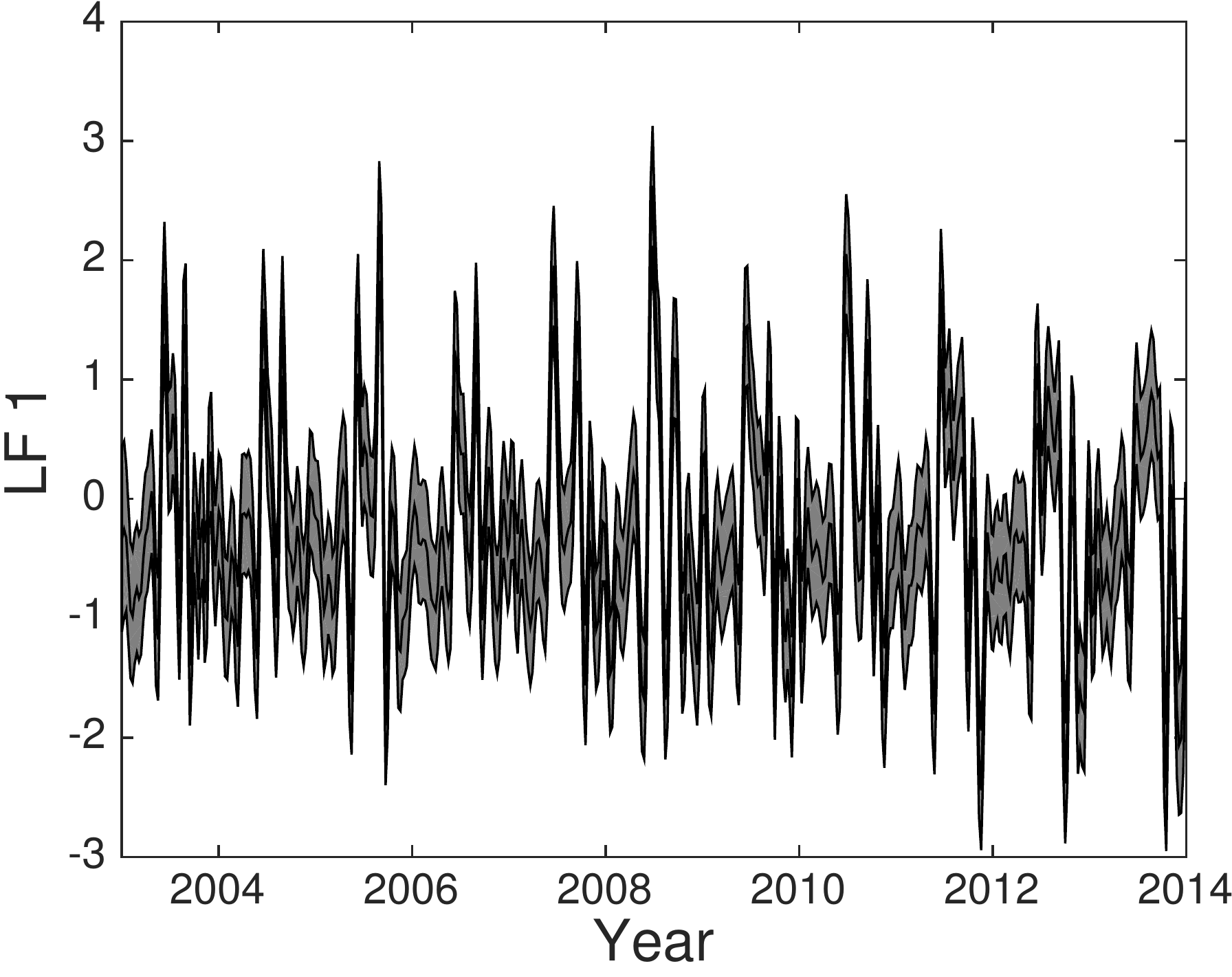}
	\includegraphics[width=0.32\textwidth]{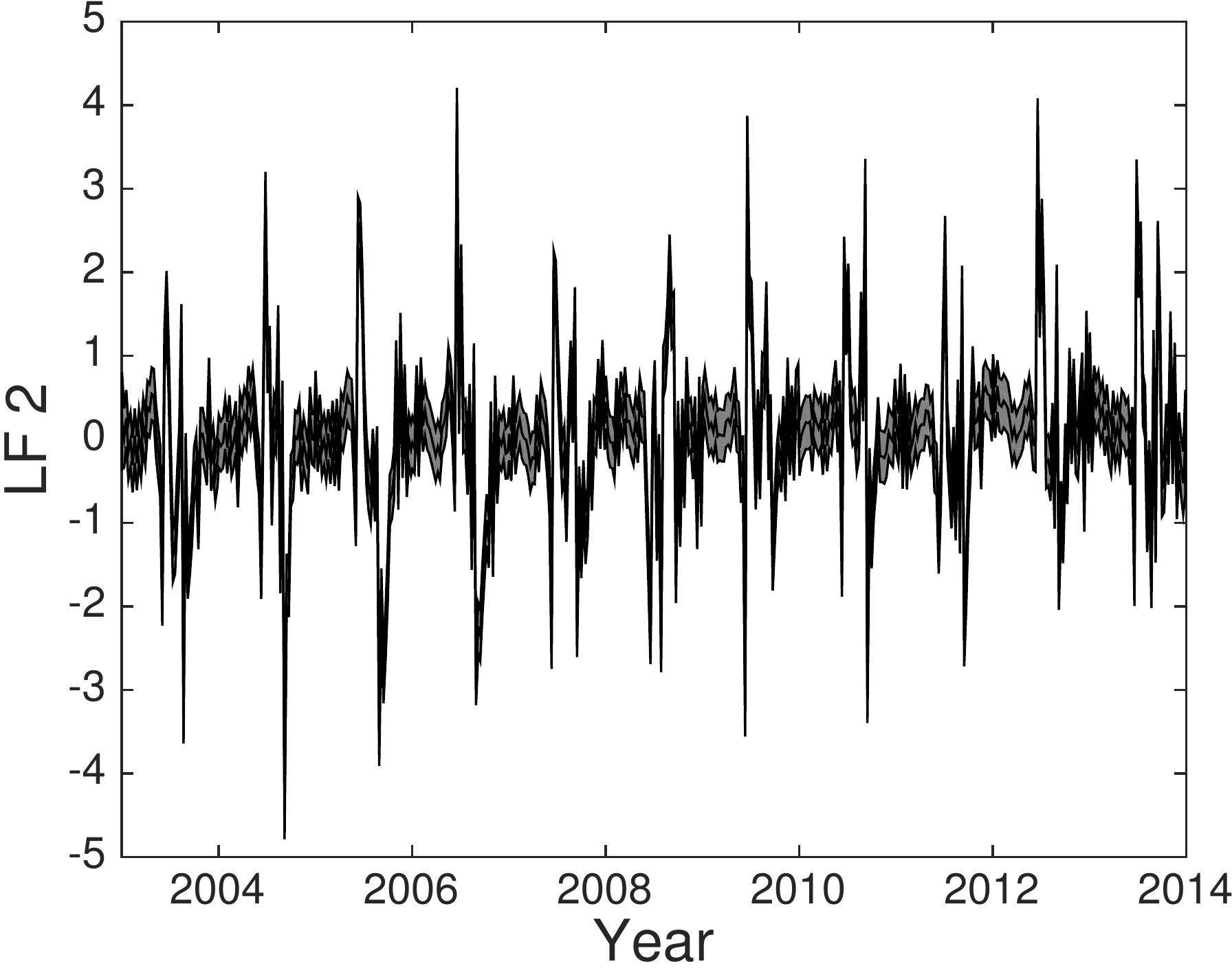}
	\includegraphics[width=0.32\textwidth]{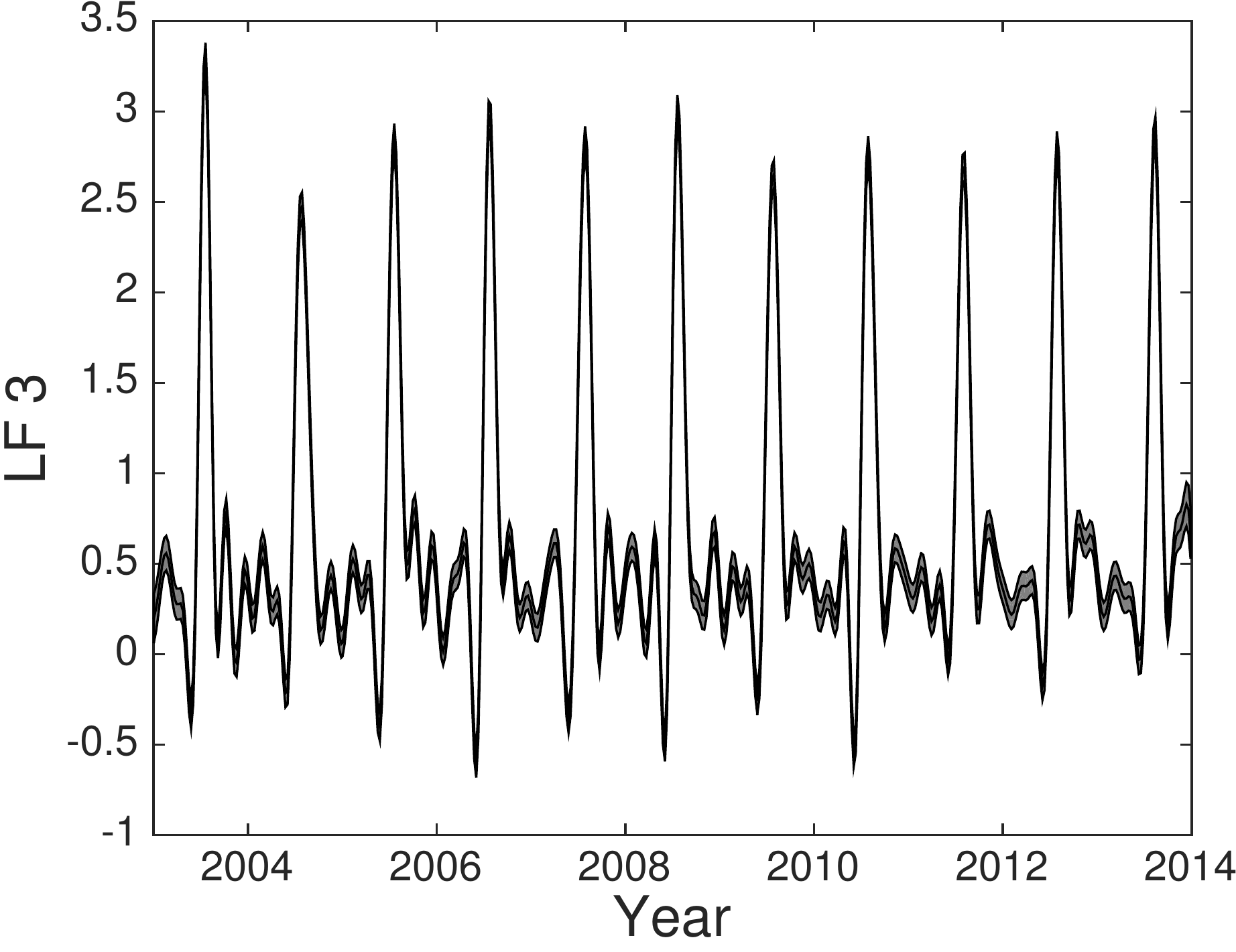}
\caption{Inferred LFs (black line) and uncertainty measured by $\pm 2$ standard deviations about the mean predicted value (grey shaded area), obtained from the full LAI and fAPAR dataset from Spain and Italy with $R=3$.}
\label{fig:LFs}
\end{figure*}

\begin{figure*}[!t]
	\centering
	\includegraphics[width=0.45\textwidth]{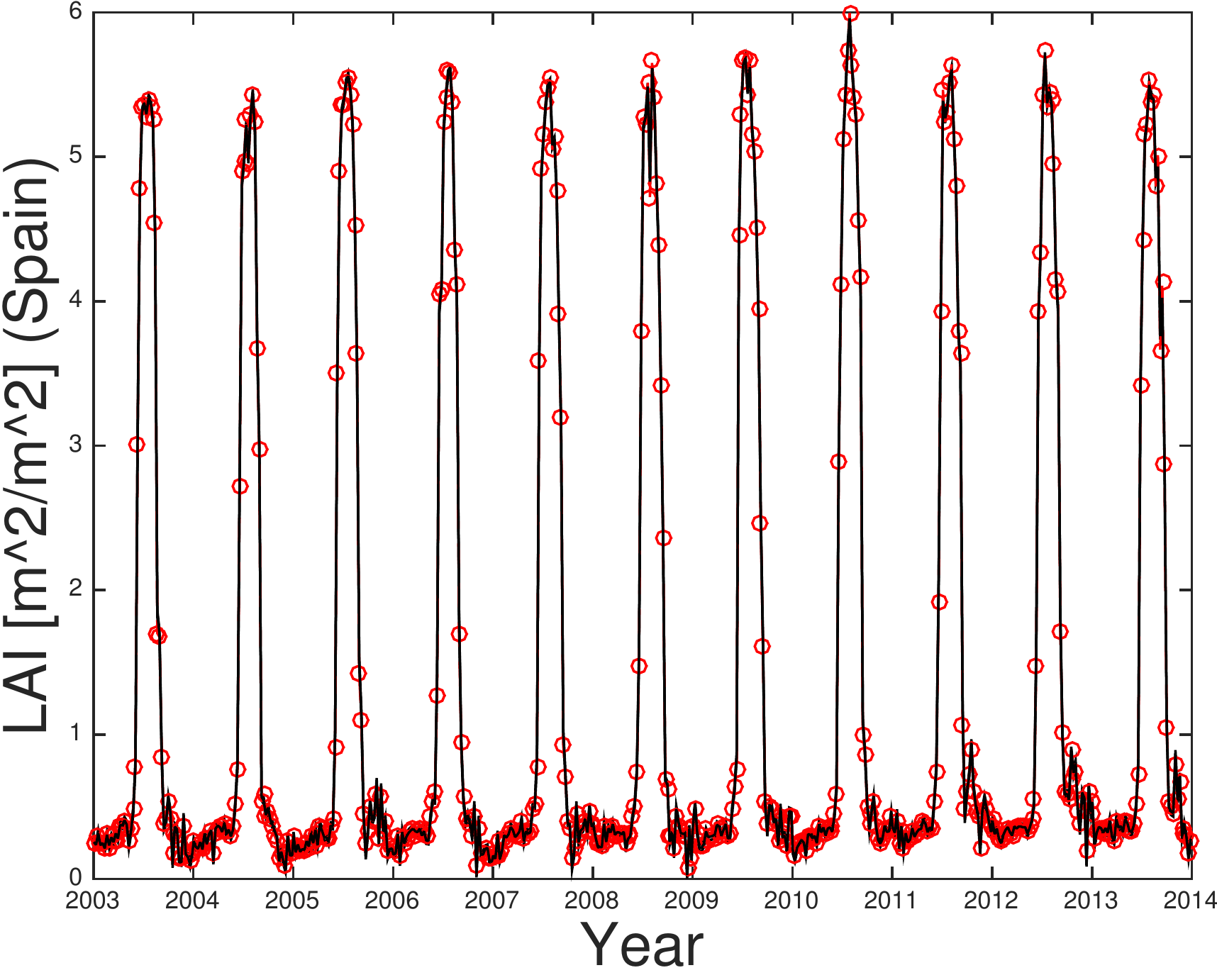}
	\includegraphics[width=0.45\textwidth]{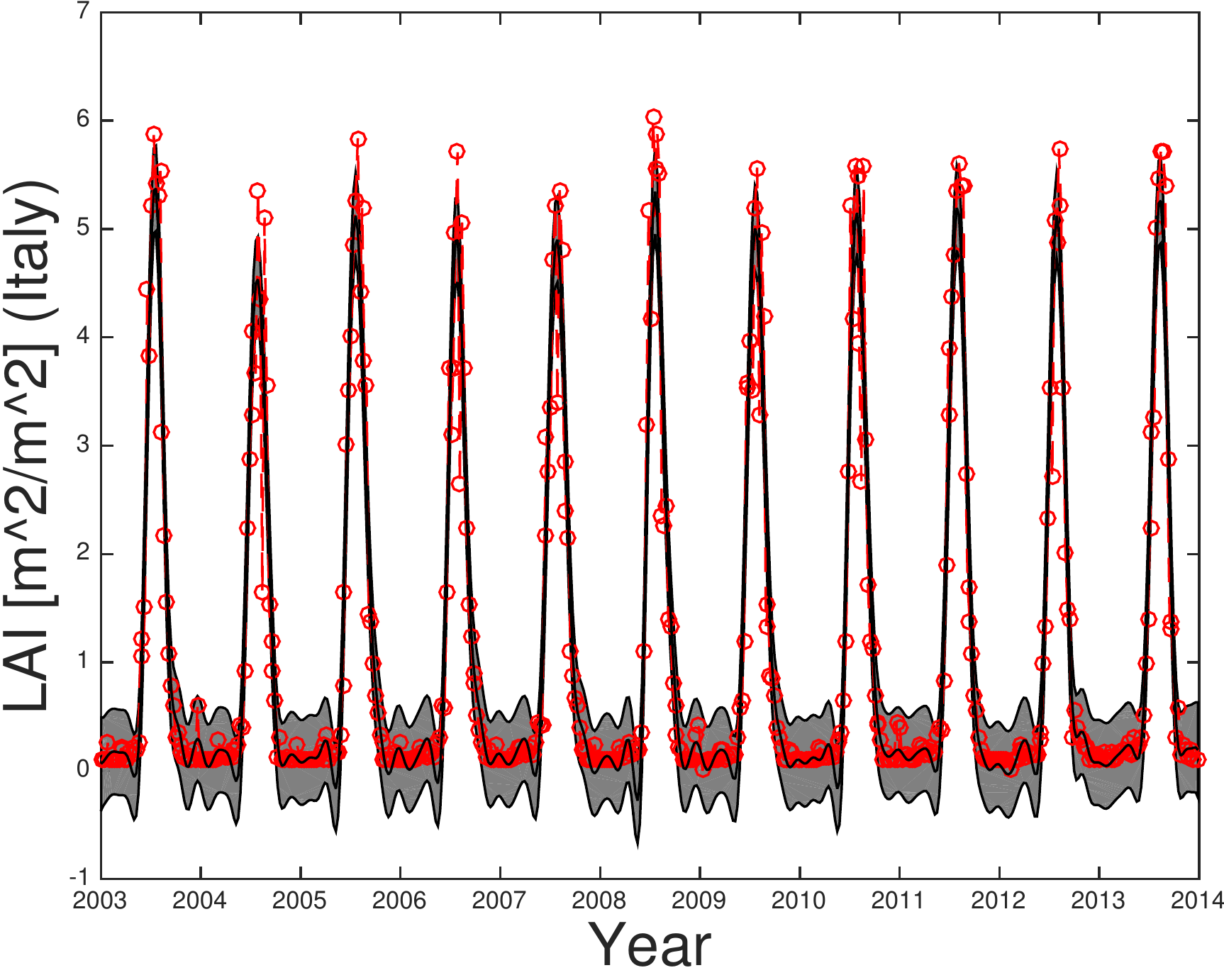}
	\includegraphics[width=0.45\textwidth]{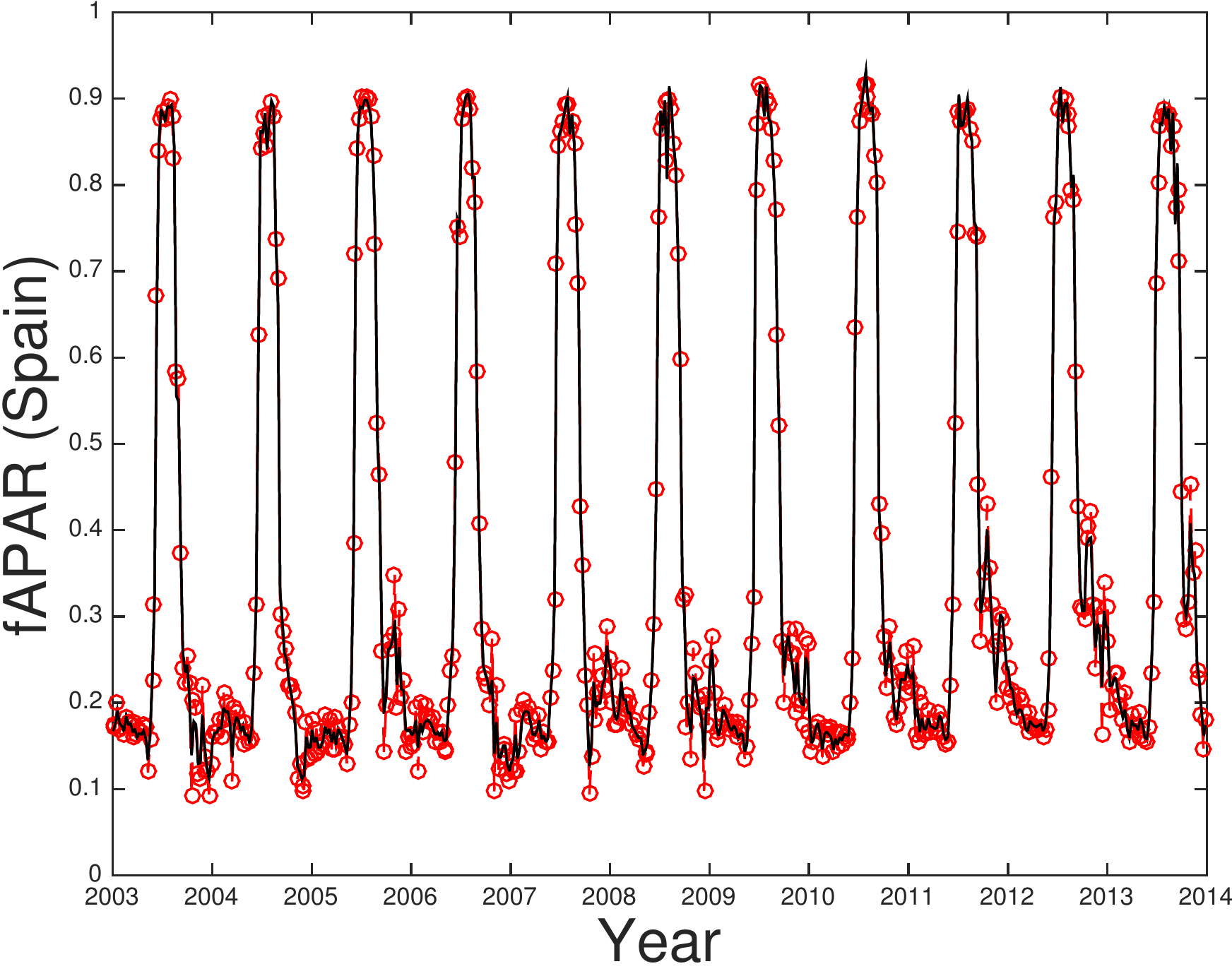}
	\includegraphics[width=0.45\textwidth]{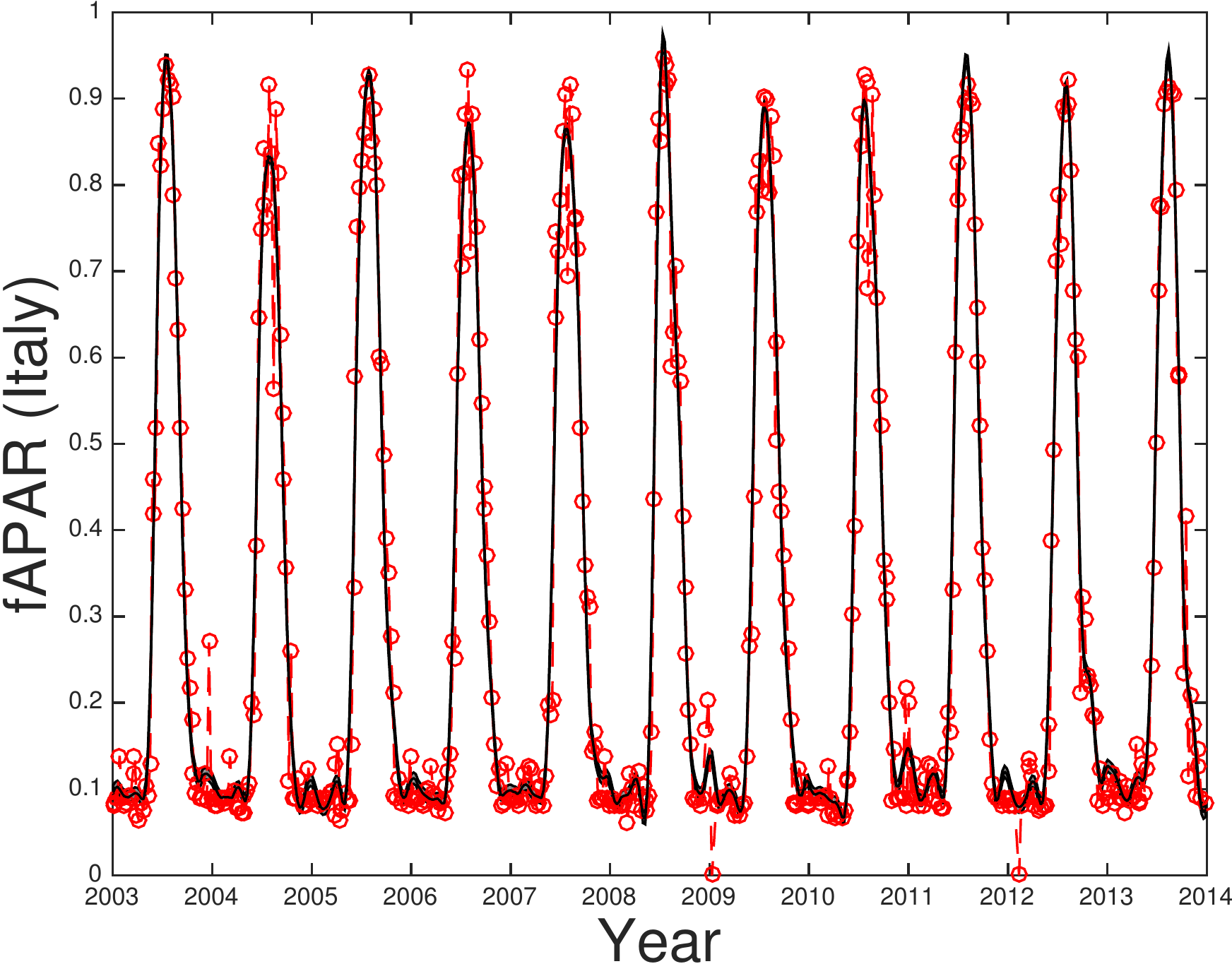}
\caption{Training data (red circles), predicted time series (black line) and uncertainty measured by $\pm 2$ standard deviations about the mean predicted value (grey shaded area), obtained from the full LAI and fAPAR dataset from Spain and Italy when $R=3$.}
\label{fig:outputs}
\end{figure*}

\begin{figure}[!tb]
	\centering
	\includegraphics[width=0.45\textwidth]{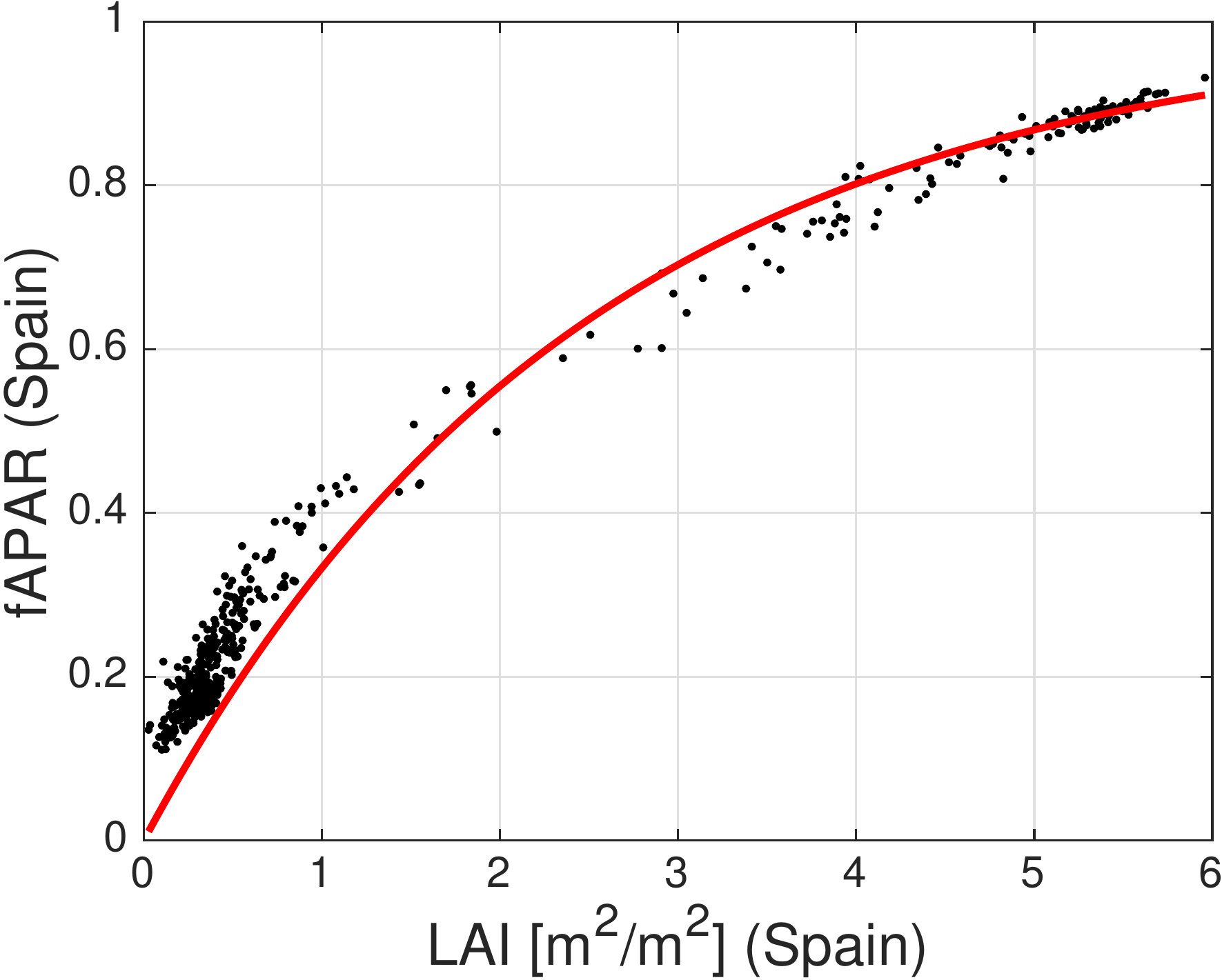}
	\includegraphics[width=0.45\textwidth]{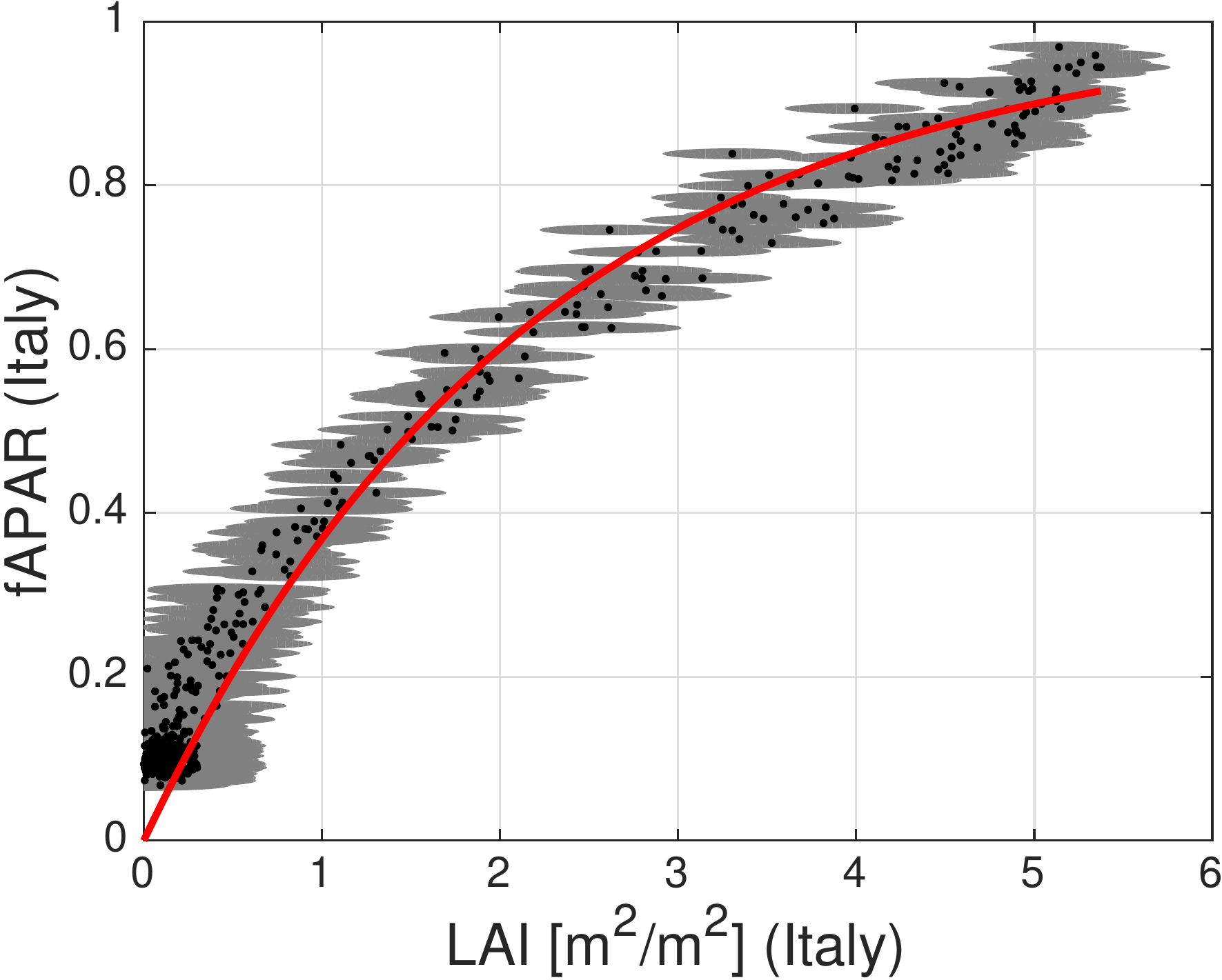}
\caption{LAI vs. fAPAR for the data learned using $R=3$ LFs and all the available data for years 2003--2013.}
\label{fig:res_LaiFapar}
\end{figure}

Noting the substantial decrease in MSE, in the rest of this section we set $R=3$.
Figure \ref{fig:LFs} shows the three LFs inferred and Figure \ref{fig:outputs} displays the four output time series.
In Fig. \ref{fig:outputs} we can see the good modeling accuracy for all the time series (as evidenced by the low NMSE values displayed in Table \ref{tab:mse_full}), whereas in Fig. \ref{fig:LFs} we see that LF3 captures the smooth and periodic component of the output and the other two LFs focus on the noisier part (albeit with an important residual periodical component).
Finally, Figure \ref{fig:res_LaiFapar} displays the LAI vs. fAPAR scatter plot, obtained from the modeled time series both for Spain and Italy.
The shaded area corresponds to the uncertainty (that appears now in both axis, as a consequence of the modeling uncertainty in both LAI and fAPAR), whereas the continuous red line shows the exponential model in \eqref{eq:fapar_vs_lai}, that has been fitted to the data by performing a simple least squares regression in the log-domain.
Note the good fit in both cases of the expected LAI vs. fAPAR relationship, given by \eqref{eq:fapar_vs_lai} with $\alpha = -0.4047$ and $\alpha = -0.4593$ for Spain and Italy respectively, and the scatter plot obtained from the GP-LFM.

\subsubsection{Dealing with missing data}

\begin{figure}[!t]
	\centering
	\includegraphics[width=0.32\textwidth]{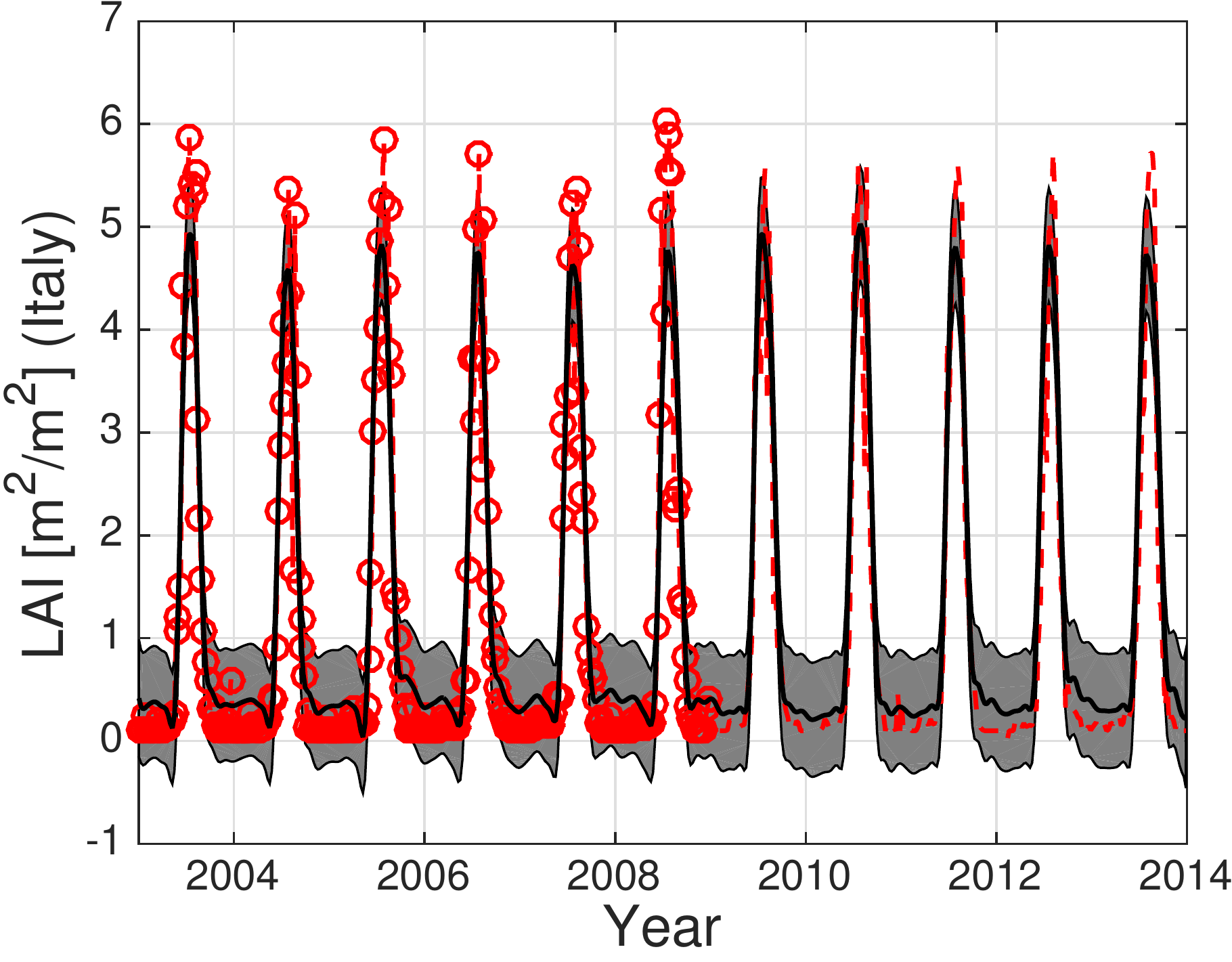}
	\includegraphics[width=0.32\textwidth]{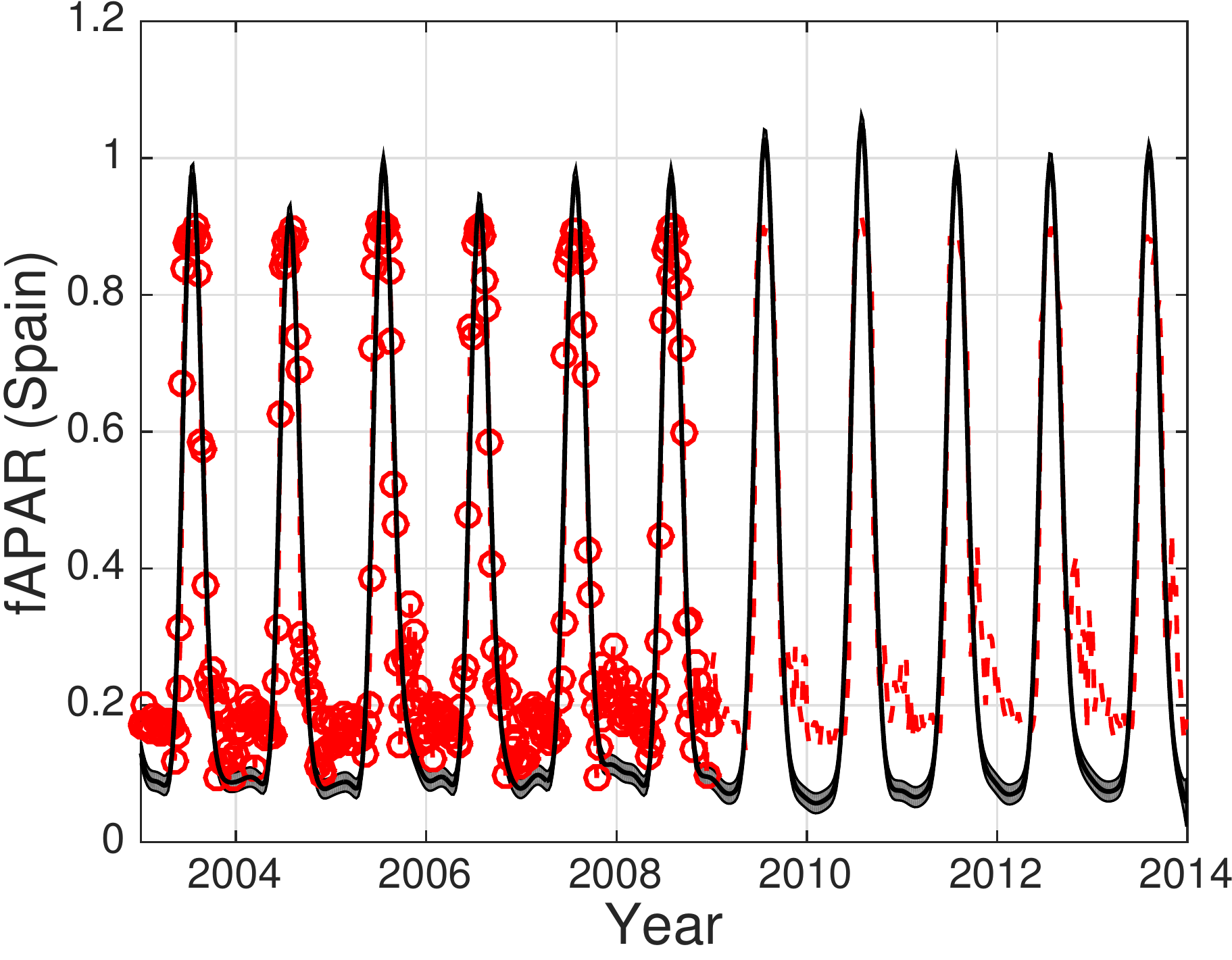}
	\includegraphics[width=0.32\textwidth]{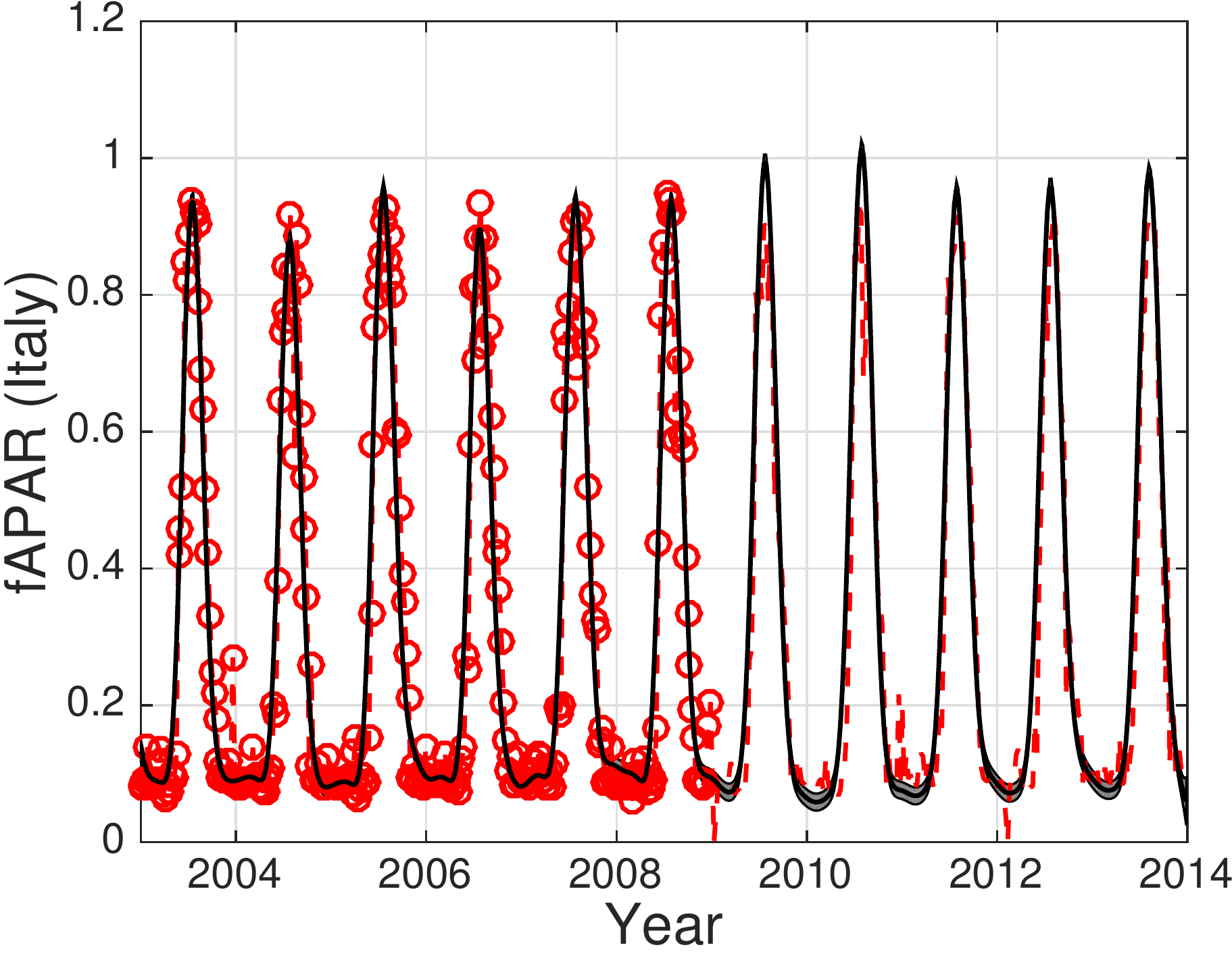}
\caption{Gap filling example using a single LF (i.e., $R=1$). Training used all the LAI data from Spain (years 2003--2013) and the first half (years 2003--2009) of the other three time series: LAI (IT), fAPAR (ES) and fAPAR (IT). The second half constitutes the test set of such time series. Training data (red circles), test data (red dashed line), predicted time series (black line) and uncertainty measured by $\pm 2$ standard deviations about the mean predicted value (gray shaded area).}
\label{fig:res_missing}
\end{figure}

In this second example, we show the ability of the model to recover the missing samples (i.e., to perform gap filling) that typically appear in this kind of datasets.
We use again all the LAI data from the MODIS sensor for Spain from the beginning of 2003 until the end of 2013 (i.e., $N_1=506$), but only the first half (years 2003--2009) of the LAI data from Italy and the two fAPAR time series (i.e., $N_2=275$, $N_3=276$ and $N_4=275$).
In order to show that the GP-LFM is able to capture the underlying dynamics of the multi-output time series with a single LF, we set $R=1$.
The three time series with missing data are displayed in Fig. \ref{fig:res_missing}, whereas the single LF (not shown) is very similar to the smooth LF (LF3) in Fig. \ref{fig:LFs}.
Note the good fit of the three time series, even though the last four years of data are removed from all of them.

\section{Automatic Emulation}\label{sec:5}

Emulation deals with the challenging problem of building statistical models for complex physical RTMs. The emulators are also called {\em surrogate} or {\em proxy} models, and try to learn from data the equations encoded in the RTM. Namely, an emulator is a statistical model {that} tries to reproduce the behavior of a deterministic and often very costly physical model. Emulators built with GPs are gaining popularity in remote sensing and geosciences, since they allow efficient data processing and sensitivity analysis~\cite{Busby09,Rivera2015,CampsValls16grsm}. Emulators also allow model tractability, as model-data integration, fast inference, analytical Jacobian calculation, and derivation of confidence intervals for the estimates and parameters becomes easier and analytical when GPs are used.

Here, we are interested in optimizing emulators such that a minimal number of simulations is run (for a given approximation error). We describe a general framework, called {\it Automatic Emulation} (AE) technique which is related to Bayesian optimization and active learning techniques. We first define the generic elements of the AE methodology and then describe a  specific implementation based on GPs. This yields the Automatic Gaussian Process Emulator (AGAPE) model for automatic emulation and creation of a compact and informative look-up-table.

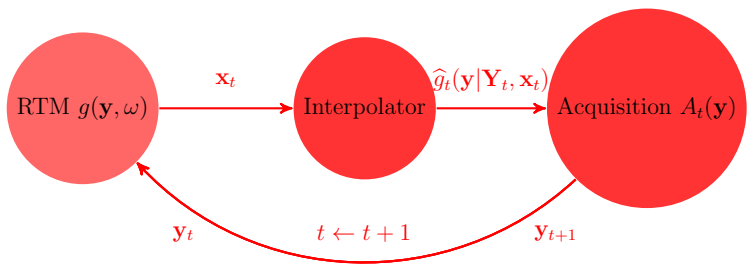
\begin{figure}[t!]
\begin{center}
\begin{tikzpicture}[->,>=stealth',scale=.75,transform shape,node distance=5cm,thick]
  \tikzstyle{every state}=[fill=red!30,draw=none,text=black]
  \node[state] (RTM)  [fill=red!60] {RTM $g(\y,\omega)$};
  \node[state] (INT)  [right of=RTM,fill=red!80] {Interpolator};
  \node[state] (ADQ)  [right of=INT,fill=red!80] {Acquisition $A_t(\y)$};
  \path[solid, thick, color=red] (RTM) edge node[yshift=+0.5cm] {$\x_t$} (INT);
  \path[solid, thick, color=red] (INT) edge node[yshift=+0.5cm] {$\widehat g_t(\y|\Y_t,\x_t)$} (ADQ);
  \path[solid, thick, bend left=45, color=red] (ADQ) edge node[xshift=-3cm,yshift=+0.5cm] {$\y_t$} (RTM);
  \path[solid, thick, bend left=45, color=red] (ADQ) edge node[xshift=+3.6cm,yshift=+0.5cm] {$\y_{t+1}$} (RTM);
  \path[solid, thick, bend left=45, color=red] (ADQ) edge node[xshift=+0.2cm,yshift=+0.5cm] {$t\leftarrow t+1$} (RTM);
\end{tikzpicture}
\end{center}
\vspace{-1cm}
\caption{Scheme of an automatic emulator.
{The goal of the model is to emulate the RTM as best as possible using a minimum number of runs. This is achieved by an iterative process which starts with a reduced input set of variables $\Y_{t=0}$. Then the RTM provides the corresponding observations $\x_{t=0}$. An interpolator is then fitted, forming part of the acquisition function that is optimized to select new variables to add to the initial input set. The process is iterated until the stop condition is fulfilled}}

\label{fig1}
\end{figure}

\begin{figure}[!ht]
	\centering
	\includegraphics[width=0.8\textwidth]{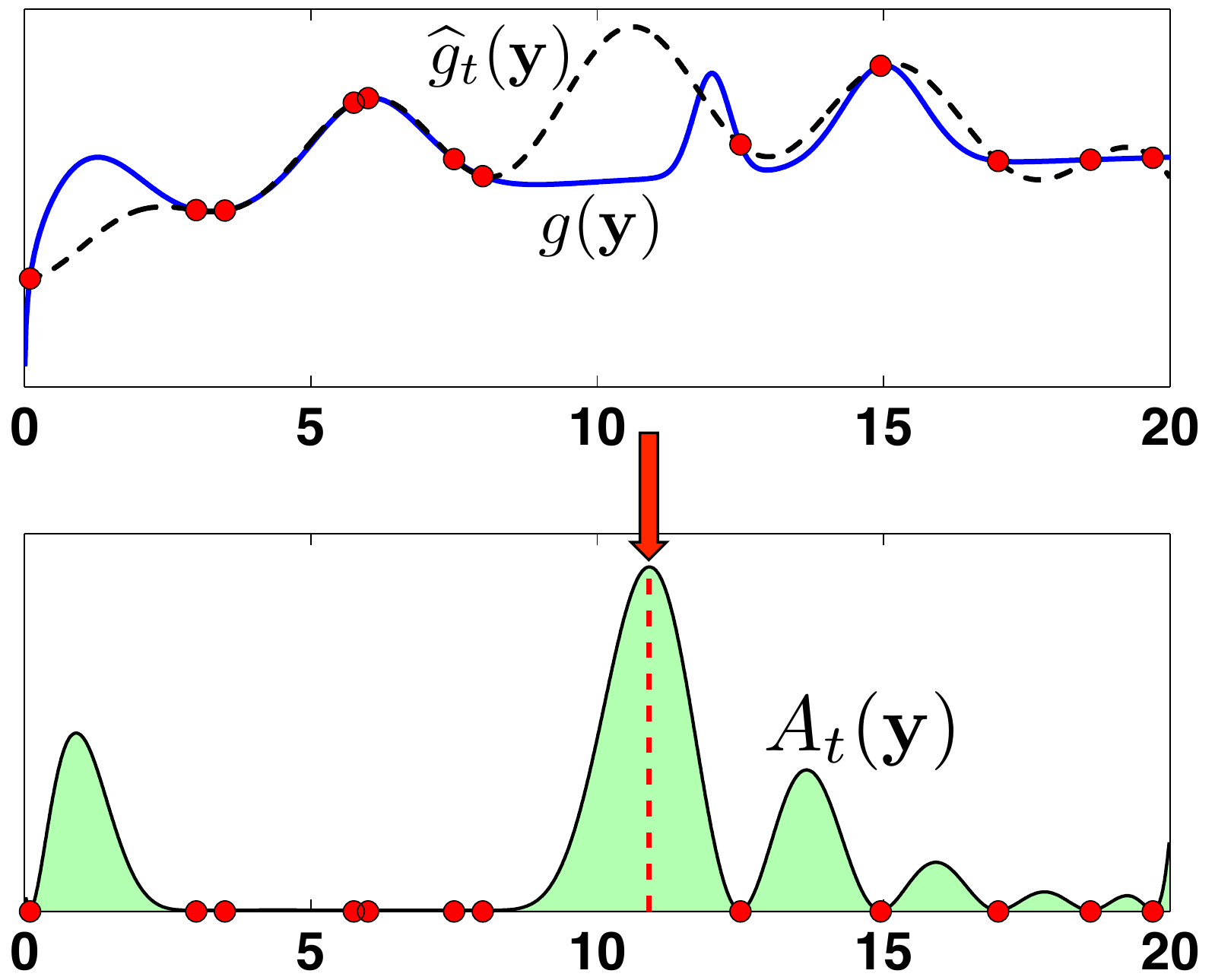}
\caption{General sketch of an Automatic Emulation (AE) procedure. The RTM model $g({\bf y})$ (top - solid line), its approximation $\widehat{g}_t({\bf y})$ (top - dashed line) and an acquisition function $A_t({\bf y})$. Its maximum suggests where adding a new node/point to the LUT.}
\label{fig:AGAPE_1}
\end{figure}

The goal is to emulate (i.e., interpolate/mimic) a costly function $g(\y)$ choosing adequately the nodes, in order to reduce the error in the interpolation with the smallest possible number of evaluation of $g(\y)$. Given an input matrix of nodes (used for the interpolation) at the $t$-th iterations, $\Y_t=[\y_1\cdots \y_{m_t}]$, of dimension $d\times m_t$ (where $d$ is the dimension of each $\y_i$ and $m_t$ is the number of points), we have a vector of outputs, ${\bf x}_t=[x_1,\ldots,x_{m_t}]^{\intercal}$, where $x_t=g(\y_t)$ is the estimation of the observations (e.g., radiances) at iteration $t\in \mathbb{N}^+$ of the algorithm. Figures~\ref{fig1}-\ref{fig:AGAPE_1} show two graphical representations of a generic automatic emulator. At each iteration $t$ one performs an {\em interpolation/regression}, obtaining $\widehat{g}_t(\y|\Y_t,\x_t)$, followed by an {\em optimization} step that updates the acquisition function, $A_{t}(\y)$, updates the set $\Y_{t+1} =[\Y_t, \y_{m_t+1}]$ adding a new node, set $m_t \leftarrow m_t +1$ and $t\leftarrow t+1$.
{Note that the acquisition function $A_t(\y)$ is the core of the automatic emulation method. It plays the role of an oracle, suggesting in which regions it is more convenient to introduce new nodes, used in the next interpolation step. Clearly, the design of $A_t(\y)$ is a key point for the success of the automatic emulation.}
The procedure is repeated until a suitable stopping condition is met, such as a certain maximum number of points is included or a desired precision error $\epsilon$ is achieved. Since $g$ is unknown, we could compute $\|\widehat{g}_t(\y)-\widehat{g}_{t-1}(\y)\|\leq \epsilon$.

\subsection{Theoretical Formulation} 
The acquisition function, $A_t(\y)$,  encodes useful information for proposing new nodes to build the emulator. {Namely, the acquisition function $A_t(y)$  suggests where introducing new support points for the next interpolation step. For this reason, the best new possible node, according to the designed acquisition function $A_t(y)$, is exactly the $\arg \max A_t(y)$.} Therefore, at each iteration,  a new node is added maximizing $A_t(\y)$, i.e.,
$$
\y_{m_{t}+1}=\arg\max A_t(\y),
$$
and set $\Y_{t+1}=[\Y_t,\y_{m_{t}+1}]$, $m_{t+1}=m_t+1$. {Observe that the acquisition function $A_t(\y)$ must take into account the locations of the current nodes and the geometry of the underlying function $g$. Hence, we propose to factorize the acquisition function as product of  a {\em geometry} $H_t(\y)$ and a {\em diversity} $D_t(\y)$ terms. The distribution of the previous nodes is encoded into the function $D_t(\y)$, whereas the geometric information is included in $H_t(\y)$.} More specifically, we define the  acquisition function as
\begin{eqnarray}
\label{AF}
A_{t}(\y)&=&\left[H_t(\y)\right]^{\beta_t}  D_t(\y), \quad \beta_t\in[0,1],
\end{eqnarray}
where $A_{t}(\y): \mathcal{Y}\mapsto\mathbb{R}$, and $\beta_t$ is an increasing function with respect to $t$, with $\textstyle\lim_{t\to \infty} \beta_t=1$ (or $\beta_t=1$ for $t>t'$). Function $H_t(\y)$ captures the geometrical information in $g$, while function $D_t(\y)$ depends on the distribution of the points in the current vector $\Y_t$. More specifically, $D_t(\y)$ presents a greater value around empty areas within $\mathcal{Y}$, whereas $D_t(\y)$ will be approximately zero close to the support points and exactly zero at the support points, i.e., $D_t(\y_i)=0$, for $i=1,\ldots,m_t$ and $\forall t\in \mathbb{N}$. Since $g$ is unknown, the function $H_t(\y)$ can be only derived from information acquired in advance or by considering the approximation $\widehat{g}$. The tempering value, $\beta_t$, helps to downweight the likely less informative estimates in the very first iterations. If $\beta_t=0$, we disregard $H_t(\y)$ and $A_{t}(\y)= D_t(\y)$, whereas, if $\beta_t=1$, we have $A_{t}(\y)=H_t(\y) D_t(\y)$.

\subsection{Specific Implementations} 

An automatic emulation (AE) procedure above described is completely defined by the following elements:
\begin{enumerate}
\item the  choice of the interpolator providing the approximation $\widehat{g}_t(\y|\Y_t,\x_t)$,  
\item the choice of the function $D_t(\y)$,
\item the choice of the function $H_t(\y)$,
\item and the choice of the tempering function $\beta_t$.
\end{enumerate}
Furthermore, the stopping condition can be considered as an additional element. For the interpolation, we have to take into account the ability of the interpolator of building the approximation in high dimensional spaces and the differentiability of $\widehat{g}_t$  (in the support domain with the exception of the a set of null measure). 
The conceptual set of  elements $\{\widehat{g}_t,D_t,H_t,\beta_t\}$  defines an AE method. Different combinations of these elements produces different AE techniques, each one yielding different performance. 

\subsection{Automatic Gaussian Process Emulator (AGAPE)} 
We consider a GP technique as interpolator with $\y_t$ as inputs and $\x_t$ as outputs (i.e., reversed with respect to the previous section). In addition, note that interpolation forces to zero the noise standard deviation, i.e., $\sigma_e=0$. Therefore, the AGAPE predictive mean and variance at iteration $t$ for a new point $\y_*$ becomes simply
\begin{align*}
\widehat g_t(\y_*) &= \bk_{*}^\intercal \bK^{-1}\x = \bk_{*}^\intercal\balpha, \\
\sigma^2(\y_*) &= k(\y_*,\y_*) - \bk_{*}^\intercal \bK^{-1}\bk_{*},
\end{align*}
where now $\bk_{*} = [k(\y_*,\y_1), \ldots, k(\y_*,\y_{m_t})]^\intercal$ contains the similarities between the input point $\x_*$ and the observed ones at iteration $t$, $\bK$ is an $m_t\times m_t$ kernel matrix with entries $\bK_{i,j}:=k(\y_i,\y_j)$, and $\balpha=\bK_{nn}^{-1}\x_t$ is the coefficient vector for interpolation. The interpolation for $\y_*$ can be simply expressed as a linear combination of $\widehat g_t(\y_*) = {\bf k}_*^\intercal{\bm \alpha} = \sum_{i=1}^{m_t} \alpha_i k(\y_*,\y_i)$. We consider the standard squared exponential kernel function now working with state vectors $\y$, i.e.  $k(\y,\y')=\exp(-\|\y-\y'\|^2/(2\delta^2))$. The training of the hyperparameter $\delta$ of kernel function $k$ can be performed with standard procedure, as cross-validation or marginal likelihood optimization \cite{OMCMC}.

Note that $\sigma^2(\y_i)=0$ for all $i=1,\ldots, m_t$ and $\sigma^2(\y)$ depends on the distance among the support points $\y_t$, and the chosen kernel function $k$ and associated hyper-parameter $\sigma$. For this reason, the function $\sigma^2(\y)$ is a good candidate to represent the distribution of the $\y_t$'s since it is zero at each $\y_i$, and higher far from the points $\y_i$'s. Moreover, $\sigma^2(\y)$ takes into account the information of the GP interpolator. Therefore, we consider as the diversity term 
$$
D(\y):=\sigma^2(\y),
$$
i.e., $D(\y)$ is induced by the GP interpolator. As geometric information, we consider enforcing flatness on the interpolation function, and thus aim to minimize the norm of the  the gradient of the interpolating function $\widehat{g}_t$ w.r.t. the input data $\y$, i.e., 
$$
H(\y) = \|\nabla_y\widehat{g}_t(\y|\Y_t,\x_t)\| = \bigg\|\sum_{i=1}^{m_t} \alpha_i\nabla_y k(\y,\y_i)\bigg\|.
$$ 
This intuitively makes wavy regions of $\widehat{g}_t$ require more support points than flat regions. The gradient vector for the squared exponential kernel $k(\y,\y')=\exp(-\|\y-\y'\|^2/(2\delta^2))$ with $\y=[y_1,\ldots,y_d]^{\intercal}$, can be computed in closed-form, $\nabla_y k(\y,\y')=-\frac{k(\y,\y')}{\delta^2}[(y_1-y'_1),\ldots,(y_d-y'_d)]^{\intercal}$. Therefore, the acquisition function can be readily obtained by defining $\beta_t=1-\exp(-\gamma t)$, where $\gamma\geq 0$, and 
$ A_{t}(\y)=\left[H_t(\y)\right]^{\beta_t}  D_t(\y)$. We optimized $A_t(\x)$ using interacting parallel simulated annealing methods,for the sake of simplicity~\cite{OMCMC,Read13}.
{Indeed, these techniques do not require the knowledge of the gradient of the acquisition function, and thus more sophisticated schemes can be employed. In our experiments, simulated annealing schemes provide good performance, reaching a solution close to the global maximum in few iterations \cite{MarParChain0,MarParChain}. However, as the dimension of the input space grows, the performance becomes worse.}    

\subsection{Multi-output Automatic Emulation}
So far we have considered an RTM model of type
$x=g({\bf y})+e$, where $e\sim \mathcal{N}(0,\sigma_e^2)$ (with $\sigma_e=0$ in AGAPE). Let us now denote the following RTM model represented by  the equation
\begin{eqnarray}
{\bf x}&=&{\bf g}({\bf y})+{\bf e},
\end{eqnarray}
with ${\bf x}=[x^{(1)},\ldots,x^{(K)}]\in \mathbb{R}^K$ and ${\bf e}\sim \mathcal{N}({\bf 0},\sigma_e^2{\bf I}_K)$ 
where ${\bf I}_K$ is a $K\times K$ identity matrix. Then, we have $K$ outputs for each ${\bf y}$, i.e.,  
\begin{equation}
{\bf g}({\bf y})=[g^{(1)}({\bf y}),\ldots,g^{(K)}({\bf y})]: \mathcal{Y} \rightarrow \mathbb{R}^{K}.
\end{equation}
In order to design an automatic emulator of $g({\bf y})$, we have to extend the strategy described above. We consider that at the $t$-th iteration the current matrix of nodes $\Y_t=[\y_1,\ldots,\y_{m_t}]$ is shared by all outputs. Therefore, we will design a multi-output emulator with a unique acquisition function $A_t(\y)$. The  multi-output emulator is completely defined by choosing the following elements:
\begin{enumerate}
\item A {\it multi-output} interpolation/regression scheme, considering the same input matrix $\Y_t=[\y_1,\ldots,\y_{m_t}]$ (for all the outputs) and the output matrix $\X_t=[\x_1,\ldots,\x_{m_t}]$, providing an approximation
$$
{\widehat {\bf g}}_t({\bf y}|\Y_t,\X_t)=[{\widehat g}^{(1)}({\bf y}|\Y_t,\X_t),\ldots,{\widehat g}^{(K)}({\bf y}|\Y_t,\X_t)].
$$ In \cite{alvarez2009latent,alvarez2010efficient,CampsValls16grsm,tuia11msvr}, different multi-output schemes are described. The simplest procedure consists in applying one independent interpolator for each output.

\item Given the matrix of current nodes $\Y_t=[\y_1,\ldots,\y_{m_t}]$ (shared by all outputs), we obtain the diversity function $D_t({\bf y})$. Generally, we can have a diversity term $D_t^{(k)}({\bf y})$ for each output (at least they can differ for hyper-parameters learned in different  interpolator/regressor), hence we can define 
\begin{equation*}
D_t({\bf y})=\prod_{k=1}^K D_t^{(k)}({\bf y}).
\end{equation*}

\item Given each ${\widehat g}^{(k)}({\bf y}|\Y_t,\X_t)$, we obtain the functions $G_t^{(k)}({\bf y})$, for $k=1,\ldots, K$.
\end{enumerate}
Therefore, we can define the complete acquisition function as 
\begin{eqnarray}
A_t(\y)&=&\left[\prod_{k=1}^K G_t^{(k)}({\bf y})\right]^{\beta_t} \prod_{k=1}^K D_t^{(k)}({\bf y}).
\end{eqnarray}
Optimizing $A_t(\y)$, we find the next node $\y_{m_{t+1}}$ to incorporate in the next iteration, $\Y_{t+1}=[\Y_t,\y_{m_{t+1}}]$. Other automatic emulator can be designed considering multiple acquisition functions $A_t^{(k)}(\y)$, one for each output.

\subsection{Emulation of Costly Radiative Transfer Codes}

We show empirical evidence of performance on the optimization of selected points for a complex and computationally expensive RTM: the MODTRAN5-based LUT. MODTRAN5 is considered as the {\em de facto} standard atmospheric RTM for atmospheric correction applications~\cite{Berk2006}. 
In our test application, and for the sake of simplicity, we have considered $d=2$ with the Aerosol Optical Thickness at 550 nm ($\tau$) and ground elevation ($h$) as key input parameters. The underlying function $g(\y)$ consists therefore on the execution of MODTRAN5 at given values of $\tau$ and $h$ and wavelength of 760 nm.
The input parameter space is bounded to 0.05~--~0.4 for $\tau$ and 0~--~3 km for $h$. In order to test the accuracy of the different schemes, we have evaluated $g(\y)$ at all the possible 1750 combinations of 35 values of $\tau$ and 50 values of $h$. Namely, this thin grid represents the ground-truth in this example.

\begin{table}[!h]
\small
\caption{Averaged number of nodes $m_t$.}
\centering
	\begin{tabular}{|c|c|c|}
    \hline
 {\bf Random}  & {\bf Latin Hypercube}  & {\bf AGAPE}    \\
\hline
\hline
 28.43   & 16.69 &  9.16  \\
\hline
\end{tabular}
\label{tab2_Ex2_Results}
\end{table}
We tested (a) a standard, yet suboptimal, random approach choosing points uniformly within $\mathcal{Y}=[0.05,0.4]\times[0,3]$, (b) the Latin Hypercube sampling~\cite{Busby09}, and (c) the proposed AGAPE. We start with $m_0=5$ points $\y_1=[0.05,0]^{\top}$, $\y_2=[0.05,3]^{\top}$, $\y_3=[0.4,0]^{\top}$, $\y_4=[0.4,3]^{\top}$ and $\y_5=[0.2,1.5]^{\top}$ for all the techniques. We compute the final number of nodes $m_t$ required to obtain an $\ell_2$ distance between $g$ and $\widehat{g}$ smaller than $\epsilon=0.03$, with the different methods. The results, averaged over $10^3$ runs, are shown in Table~\ref{tab2_Ex2_Results}. AGAPE requires the addition of $\approx 4$ new points to obtain a distance smaller than $0.03$.

\subsection{Example of Multi-output Emulation}

We consider a multi-output toy example with scalar inputs where we can easily compare the achieved approximation $\widehat{{\bf g}}_t(y)$ with the underlying function ${\bf g}(y)$, which is unknown in the real-world applications. In this way, we can exactly check the true accuracy of the obtained approximation using different schemes. For the sake of simplicity, we  consider the following multi-output mapping
\begin{equation}
\label{TRUE_f}
{\bf g}(y)=[\log(y),0.5\log(3y)], \quad y\in(0,10],
\end{equation}
then $d=1$ and $K=2$. Even in this simple scenario, the procedure used for selecting new points is relevant as confirmed by the results provided below. We start with $m_0=4$ support points,  $\Y_0=[0.1, 3.4, 6.7, 10]$. We apply one independent GP for each output.
We add to $\Y_t$ sequentially  $20$ additional points, using different sampling strategies:
\begin{itemize}
\item[(a)] the multi-output version of AGAPE (denoted as AMOGAPE),
\item[(b)] uniform points randomly generated in $(0,10]$,
\item[(c)] a sequential Sobol sequence, 
\item[(d)] and a sequential version of the Latin Hypercube procedure (Seq-LHC).
\end{itemize}
In this last case, i.e., for Seq-LHC, $20$ points are generated following the LHC procedure and then one of them is added to ${\bf Y}_t$ at each iteration (without replacement).  
Note that, at each run, the results can vary even for the deterministic procedure due to the optimization of the hyperparameters (we use a parallel simulated annealing approach that is a stochastic optimization technique~\cite{OMCMC,MarParChain2, kirkpatrick83}). We average all the results over $500$ independent runs.
 
\begin{figure}[htb]
\centering
\centerline{
\includegraphics[width=7cm]{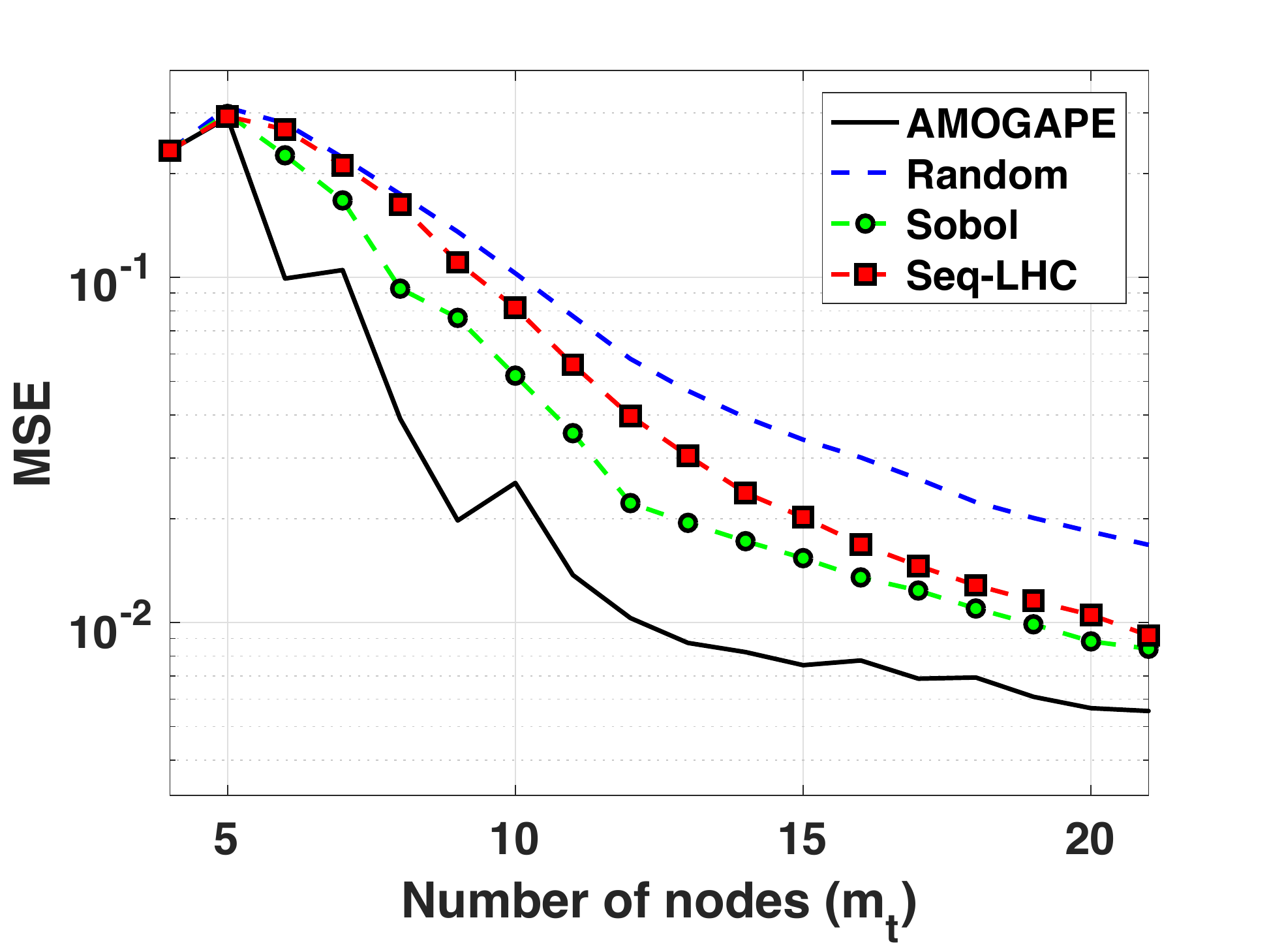}
}
\vspace{-0.25cm}
\caption{MSE (in log-scale) between ${\bf g}(y)$ and $\widehat{{\bf g}}_t(y)$ versus the number of the number of support points $m_t$, that is $m_t=t+4$ in this example. 
}
\label{FigAMOGAPE}
\end{figure}

We compute the $L_2$ distance,i.e., the Mean Square Error (MSE), between $\widehat{{\bf g}}_t(y)$ and ${\bf g}(y)$ at each iteration, obtained by the different method. We show the evolution of the averaged MSE versus the number of support points $m_t$ (that is $m_t=t+m_0$) in Figure \ref{FigAMOGAPE}. We can observe that the AMOGAPE scheme outperforms the other methods, providing the smallest MSEs between ${\bf g}(y)$ and $\widehat{{\bf g}}_t(y)$. 

\section{Conclusions}\label{sec:6}
 
This paper treated the problems of forward and inverse modeling in remote sensing using advanced machine learning methods. We presented the field formally, revised the concept of inverting or emulating radiative transfer models, and identified the main current shortcomings and trends when using machine learning in these settings. \\\\
The first section of the paper introduced the theory and use of GP models in real world applications, including retrieval of biophysical parameters for assessment and decision making in crop management and filling gaps in time series of regional EO products. GP models provide high prediction accuracy and error bars for the predictions that can be useful for masking poor predictions or detecting anomalies. The models have now been adopted or are being considered for implementation in operation processing chains. \\\\
In the following sections we payed attention to advanced GP models. All of the presented methods were motivated by observing that two (apparently contradictory) philosophies are typically adopted: either trusting the physical rules encoded in the physical model (for both simulation or inversion), or directly relying on data and thus following data science approaches (for both emulation and retrieval). We posit here that a richer, more appropriate approach to these problems emerges by developing machine learning algorithms that respect the Physics, that can incorporate prior knowledge, and that provide not only accurate but also credible predictions. 
Three types of physics-aware GP models were introduced: a simple approach to combine {\em in situ} measurements and simulated data in a single GP model, a latent force model that incorporates ordinary differential equations, and an automatic compact emulator of physical models through GPs. The developed models demonstrated good performance, adaptation to the signal characteristics and transportability to unseen situations. We analyze the main features, pros and cons of the proposed models in what follows.\\ \\
The joint GP model is a very useful approach whenever one has access to real and simulated data by a physical model. By definition, simulations can cover a large parameter space. This is why the JGP model can help to extrapolate into the regions where real data is scarce. The JGP model is actually quite conservative and tends to perform as well or better than using real data alone. Two main shortcomings were identified though: 1) the more simulated data added, the better performance is obtained in general but the training process will be slower; and 2) There needs to be simulated data in the region of the real data, otherwise the model is not going to trust that the simulated data can be effectively used to improve predictions. In our experiments we illustrated the performance of the JGP in extrapolation across sites, thus demonstrating that the model learns a sort of domain adaptation by grounding on the simulated data that should be equally relevant in different sites. \\ \\
A second approach for inverse modeling (retrieval) presented here relies on latent force models, that is, in models that encode differential equations that model the generating system well into GPs. The GP-LFM allows us to model the correlation among multiple related outputs automatically by mixing the black-box, data-driven approach characteristic of machine learning approaches with the physical information used to derive purely mechanistic models. The GP-LFM should be used when: (1) data for multiple correlated outputs is available; (2) input-output relationships in the form of differential equations are known (either exactly or approximately). The GP-LFM has the following advantages: (1) it combines the rigour in the incorporation of the available information of data-driven Bayesian approaches with the problem description accuracy of purely mechanistic models; (2) it can be cast as a generative model where the input-output relationships are not enforced directly (as in other multi-output/task approaches), but through physically interpretable latent forces (GPs) that are automatically learned from data; (3) some physical interpretability can be gained from the learned hyperparameters of the latent GPs and the smoothing kernels used to model input-output relationships; (4) the LFM-GP can be directly applied to outputs with very different characteristics, dimensions and sampling rates; and (5) the model showed very good extrapolation capabilities, thus being able to deal with missing data and to make predictions in regions not covered by the available data. Some disadvantages can be identified though: (1) the model needs to work out the expressions of the output kernels for each combination of input kernels and smoothing kernels; (2) high computational cost of GPs, which is increased by the fact that output kernels are more complex than standard kernels used in GP regression; (3) cannot incorporate non-linear differential equations, but only their linearized versions, as the whole input-output process would not be a GP any more.\\
The last novel methodology presented here has to do with the emulation of costly RTMs. For this we presented AGAPE, an automatic sequential interpolator/regressor that iteratively selects the parameter region to sample from, and build an optimal (compact) emulator. Additional advantages of AGAPE are that provides the information where possible new nodes should be incorporated to better approximate the unknown, underlying function encoded in the RTM. The performance of AGAPE depends on a good choice of the starting points (and, as consequence, of an enough number of them), and a well-designed tempering function. These two considerations are connected: indeed, the key point is how much we believe/trust the estimation of the gradient from the previous function approximation. For the same reasons, as the dimensionality of the problem grows, AGAPE requires a greater number of starting nodes, in order to ensure the reliability of the initial approximation of the unknown function.  \\\\
The framework that we presented here for attaining Physics-aware machine learning was illustrated using Gaussian Processes because of the solid grounds on Bayesian inference, properties and mathematical tractability. However, it has not escaped our notice that other machine learning algorithms could be equally applied. These issues will be subject of future research.

\section*{Acknowledgments}

The research was funded by the European Research Council (ERC) under the ERC-CoG-2014 SEDAL project (grant agreement 647423), and the Spanish Ministry of Economy and Competitiveness (MINECO) and the European Fund for Regional Development (ERDF) through the project TIN2015-64210-R. The authors would like to thank the Institute for Electromagnetic Sensing of the Environment, the Cereal Institute of DEMETER, and the Aristotle University of Thessaloniki for providing the Italian and Greek field data acquired under the ERMES FP7 project.

\end{document}